\documentclass{aa}

\usepackage{amsmath}
\usepackage{txfonts}         
\usepackage{natbib}
\usepackage{lscape}          
\usepackage[usenames]{color}

\usepackage{enumerate}
\definecolor{darkblue}{rgb}{0,0,.5}

\bibpunct{(}{)}{;}{a}{}{,}   

\title{Spectral analysis of 636 white dwarf - M star binaries from the Sloan Digital Sky Survey}

\subtitle{}

\author{R. Heller \inst{1} \and D. Homeier \inst{2} \and S. Dreizler \inst{3} \and R. \O stensen \inst{4}}

\institute{Hamburger Sternwarte (Universit\"at Hamburg), Gojenbergsweg
112, 21029 Hamburg, Germany\\
\email{rheller@hs.uni-hamburg.de}
\and
Institut f\"ur Astrophysik, Georg-August-Universit\"at
G\"ottingen, Friedrich-Hund-Platz 1. 37077 G\"ottingen, Germany\\
\email{derek@astro.physik.uni-goettingen.de}
\and
Institut f\"ur Astrophysik, Georg-August-Universit\"at
G\"ottingen, Friedrich-Hund-Platz 1. 37077 G\"ottingen, Germany\\
\email{dreizler@astro.physik.uni-goettingen.de}
\and
Institute of Astronomy, Katholieke Universiteit Leuven, Celestijnenlaan 200D, 3001 Leuven, Belgium\\
\email{roy@ster.kuleuven.be}
}

\date{Received July 17, 2008 / Accepted October 3, 2008}

\abstract{We present a catalog of 857 white dwarf (WD)-M binaries
from the sixth data release (DR6) of the Sloan Digital Sky Survey
(SDSS), most of which were previously identified. For 636 of them,
we complete a spectral analysis and derive the basic parameters of
their stellar constituents and their distances from Earth. } { We
attempt to measure fundamental parameters of these systems by
completing spectral analyses. We propose to test models typically
applied in fitting procedures and constrain likely and appropriate
evolutionary scenarios for the systems.} { We use a $\chi^2$
minimization technique to decompose each combined spectrum and
derive independent parameter estimates for its components. The
possibility of alignment by chance is demoted to statistical
insignificance, hence, we use physical interaction of the binary
constituents as input parameter. Additionally, we check the
corresponding photometric data from the SDSS to find optically
resolved systems. } { Forty-one of the stellar duets in our
spectroscopic sample are optically resolved in their respective
SDSS images. For these systems, we also derive a minimum true
spatial separation and a lower limit to their orbital periods,
typically which are some $10^{4}$\,yr. Spectra of 167 stellar
duets show significant hydrogen emission and in most cases no
additional He\,\textsc{i} or He\,\textsc{ii} features. We also
find that 20 of the 636 WDs are fitted to be DOs, with 16 measured
to have $T_{\mathrm{eff}}^{\mathrm{WD}}$ around 40\,000\,K.
Furthermore, we identify 70 very low-mass objects (VLMOs), which
are secondaries of masses smaller than about 0.1 $M_{\sun}$, to be
candidate substellar companions. } { Although various selection
effects may play a role, the fraction 6.4\,\% of WD-M star
binaries with orbital separations of around 500\,AU is a criterion
for evolutionary models of stellar binary systems. Of the 167
spectra with hydrogen emission, 8 had already been found to be
post-common envelope binaries (PCEBs) and 4 are systems with strong irradiation processes on the M dwarf. The remaining 155 Balmer-emitting binaries
probably harbor an active M dwarf (dM), corresponding to a
fraction of 24.4\,\%. The excess of cool DOs is most likely due to
additional WDs in the DB-DO $T_{\mathrm{eff}}$ range, for which no
detailed fitting was completed. The trend of the M stars being
closer to Earth than the WD component is probably due to an
underestimation of the theoretical M star radii.}

\keywords{(Stars:) binaries: spectroscopic - (Stars:) white dwarfs - Stars: late-type - Stars: fundamental parameters - Methods: data analysis - (Stars:) binaries: visual}

\begin{document}

\titlerunning{WD-M star binaries from the Sloan Digital Sky Survey}

\authorrunning{Heller et al.}

\maketitle

\section{Introduction}
\label{sec:intro}

The study of white dwarf (WD)-M star systems is an important
research area because they constitute a common final stage object
of stellar evolution, a WD, and the most frequent type of star:
the M component. Close WD-M binaries are also progenitor
candidates for cataclysmic variables (CVs) and Type Ia supernovae.
WD-M binaries consist of two stars of radically different
structure and evolutionary stage that originate in interstellar
matter (ISM) of identical composition at the same time, on
cosmological scales, and in the same region of space. Many of the
constituents of the systems do not only interact gravitationally,
which may lead to mutual mass exchange and substantial
gravitational radiation, but they also affect each other due to
their magnetic fields. With this paper, we attempt to improve the
knowledge of the basic parameters of WD-M systems.

\begin{figure*}
  \centering
  \scalebox{0.45}{\includegraphics{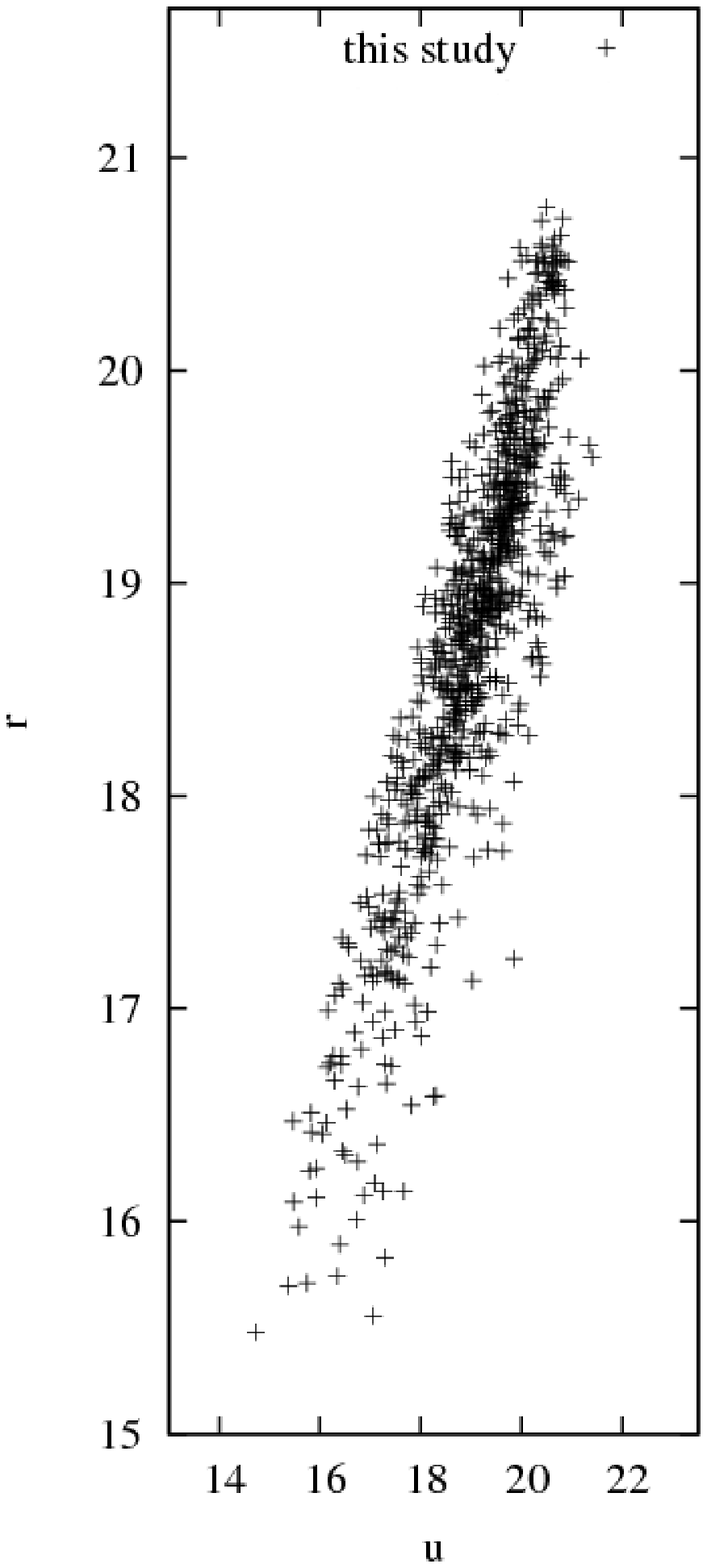}}
  \hspace{1.2cm}
  \scalebox{0.45}{\includegraphics{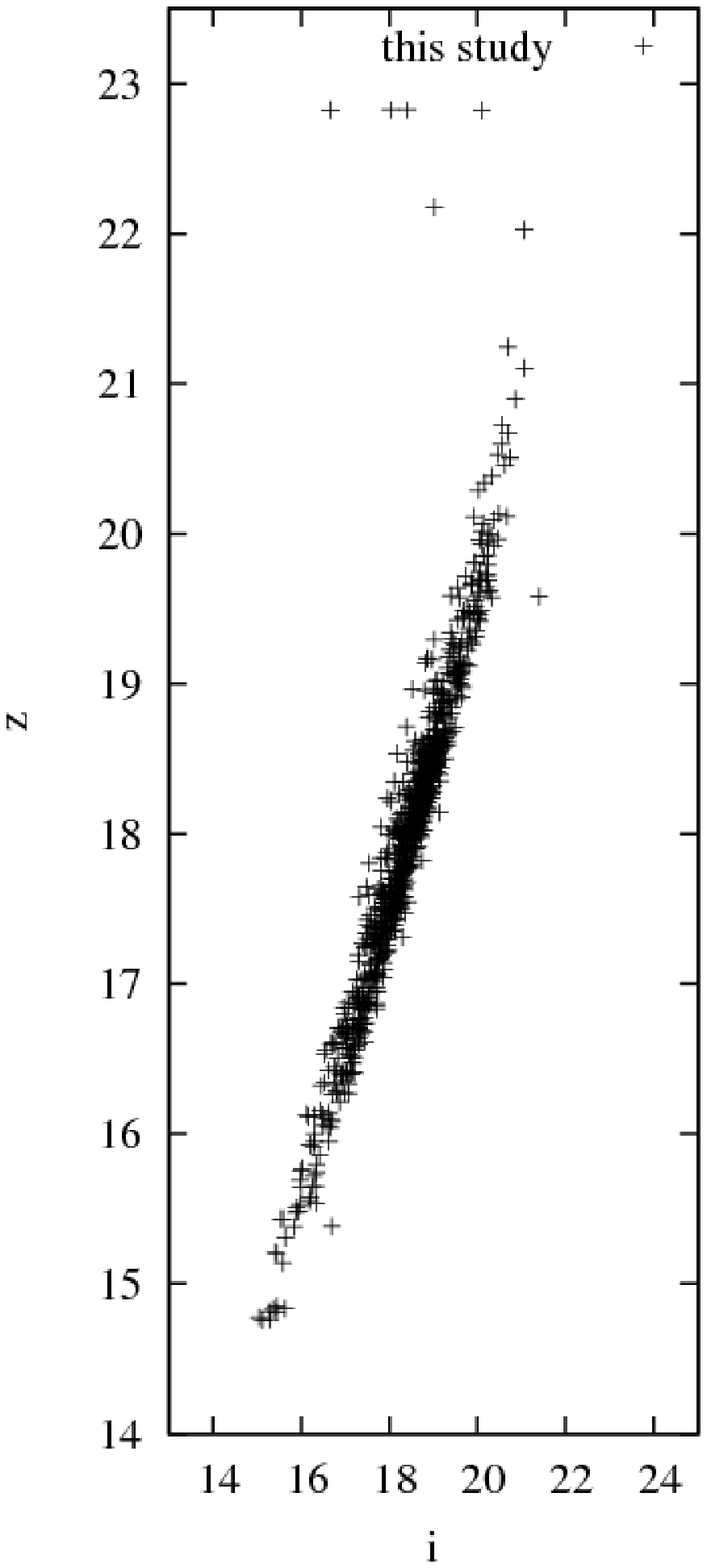}}
  \hspace{1.2cm}
  \scalebox{0.45}{\includegraphics{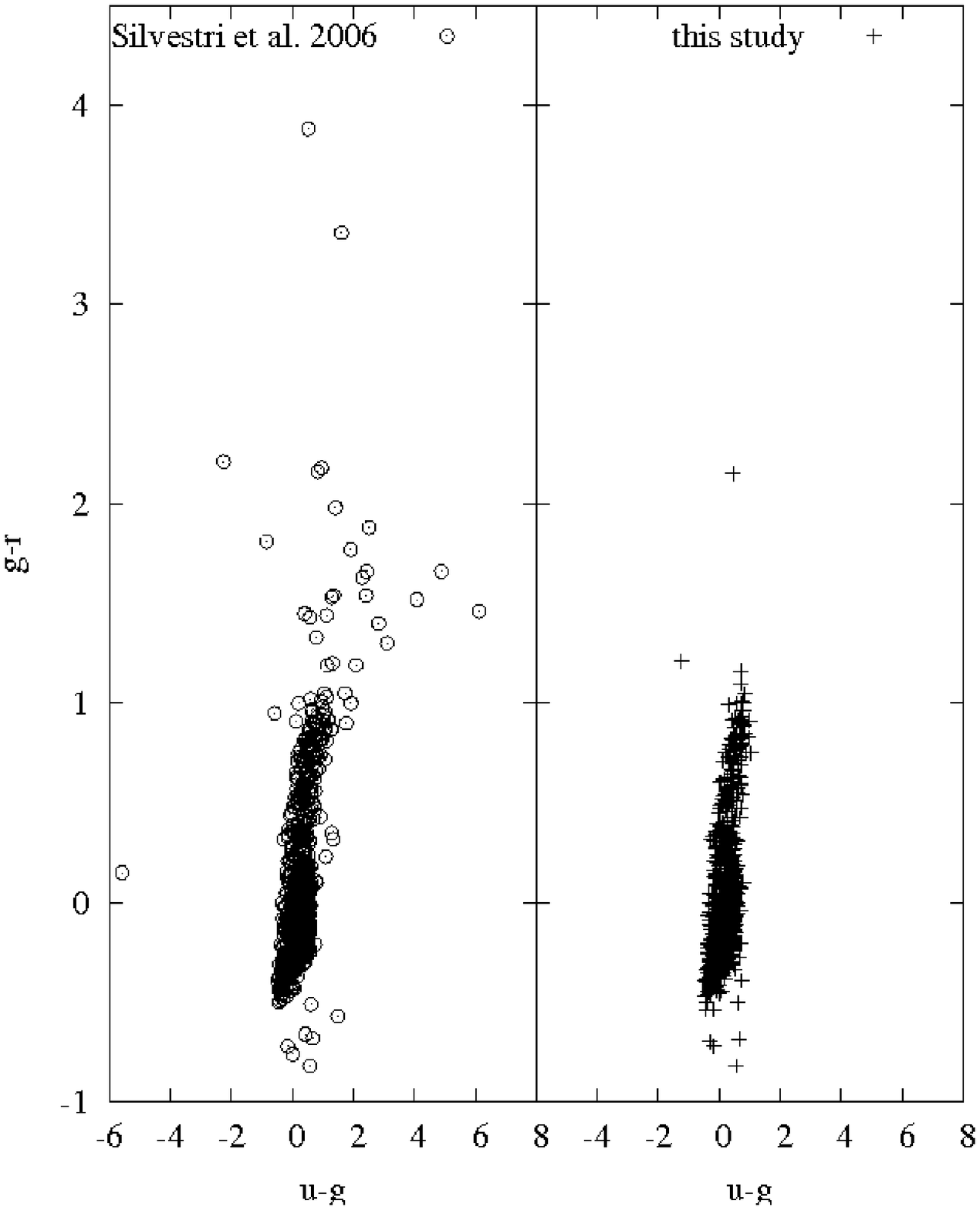}}
  \caption{\textit{Left}: The localizations of our master sample in the $(u-r)$-space; \textit{Middle}: The localizations of our master sample in the $(i-z)$-space; \textit{Right}: Comparison of the color-color localizations of our master sample with the WD-M sample of \citet{2006AJ....131.1674S}. Those objects from our study with $g-r > 0.2$ come from the EDR, on which we did not impose the ($g-r$)-cutoff.}
 \label{fig:color}
\end{figure*}

The Sloan Digital Sky Survey (SDSS) provides an invaluable observational data set with which to complete such a study. \citet{2003AJ....125.2621R} first attempted to study WD-M dwarf (dM)
pairs in the SDSS and identified 109 of these objects. In our master sample, there are 99 of them but we discarded 32 for our purposes because of reasons given at the end of Sect.
\ref{sec:Observations}. \citet{2007AJ....134..741S} presented a study
of 1253 detached close binary systems from the SDSS data release 5
(DR5) \citep{2007ApJS..172..634A}, most of them which consisted of WD primaries
and M dwarf companions. Our master sample comprises 857 WD-M star systems, 690 of which have been published
by \citet{2007AJ....134..741S}. \citet{2007MNRAS.382.1377R} presented
investigations of 101 white dwarf main-sequence binaries (WDMSs)
with multiple spectra from the SDSS, using model atmosphere
   fitting for the WD primaries and spectral typing based on the
   M-dwarf template spectra of \citet{1998A&A...339..518B} and
   \citet{2007AJ....133..531B} for the secondaries. Significant radial velocity variations were found for 18 systems, indicative of sufficiently close orbits to be post-common envelope binaries (PCEBs), and
Sect. \ref{subsec:emission} is devoted to this subject. Another study of WD-M binaries was completed by
\citet{2003ApJ...586.1356W}, who focused on the observational
identification methods with data from the Two Micron All-Sky
Survey (2MASS) incremental data release. Catalogs of WDs in the
SDSS were given by \citet[DR1]{2004ApJ...607..426K} and
\citet[DR4]{2006ApJS..167...40E} and many are included in
our sample. Four binaries from our master sample are in Luyten's WD catalogs for the years from 1970 to 1977 \citep{1999yCat.3070....0L}, which contains 6546 objects, and no binary is in his sample of 6210 objects from 1940 to 1987 \citep{1997yCat.1130....0L}. The original accuracy of the position in J1950 coordinates was 0.1\,min in Right Ascension (1.5\,$\arcmin$ or 0.025\,$^{\circ}$) and 1\,$\arcmin$ in Declination (0.017\,$^{\circ}$), as stated in the respective VizieR reference. We used the J2000 coordinates calculated by VizieR to relate stars with each other. Another earlier study of WDs in wide binaries was completed by \citet{1986AJ.....92..867G}, and we found three of his 56 systems to be in our master sample, one of a quality too low for a reliable analysis. We also show fits for the three binary systems mentioned in \citet{2006A&A...454..617H}. From our sample, 153 systems were not included in any of the mentioned studies.

The present work introduces a novel method of analyzing WD-dM
binaries by fitting simultaneously model spectra for both the white
dwarf and the main-sequence star to the composite spectrum to derive
the atmospheric parameters of both stars. Unlike the
   aforementioned studies, we thus obviate the need to calibrate
   empirical relations between physical parameters and spectral
   types. Although current cool-star atmosphere models may still have some
   shortcomings in fitting M-dwarf spectra, this approach avoids possible
   biases due to systematic differences between single M-dwarfs and those
   in close binaries.

In Sect. \ref{sec:Observations}, we provide an overview of the
observational principles and our target selection criteria. We
also state the premises for our procedure, e.g. we calculate the
probability of WD-M alignment by chance (Sect.
\ref{sub:premises}). The calculation of the theoretical spectra used
in the fitting process is described in Sect.
\ref{sec:Models}. In Sect. \ref{sec:Math}, we then provide a detailed description of
the mathematical methods we used: This represents the major part of this paper,
since the $\chi^2$ method that we use differs from
those typically applied to decompose combined spectra. With our
relatively large sample of 636 systems, we present a statistical
refinement of the SDSS WD and the M dwarf temperature functions in
Sect. \ref{sec:results}. We also find 41 optically resolved
binaries with accordingly wide-separated constituents of long
orbital periods (Sect. \ref{subsec:optical}).

\section{Observations}
\label{sec:Observations}

The observational data were taken from the SDSS DR6, which was performed using the 2.5-meter telescope at the Apache Point Observatory in southern New Mexico through June 2006 \citep{2008ApJS..175..297A}. WD-M star binaries are a byproduct of this survey which was focused on quasi-stellar objects. The most significant part of the analysis for this paper was completed using the SDSS spectra, covering the
spectral range from 3\,800 to 9\,200 $\AA$ at an inverse resolution of $\lambda/\Delta \lambda = R \approx 1800$. Targets for SDSS spectroscopic follow-up observations were selected on the basis of photometric results.

The spectroscopic dataset used in this study consisted of 857
objects that were labeled as WD-M dwarf binaries during a search
for hot subdwarf stars in the SDSS dataset. Finding hot subdwarfs
is relatively easy, because they can be clearly distinguished from
normal stars by their UV-excess, a property that they have in
common only with WDs. While the hot subdwarfs dominated the
population of UV-bright stars in earlier studies such as the PG
survey \citep{1986ApJS...61..305G}, the depth of the SDSS ensures
that the numbers of hot subdwarfs taper out as the volume surveyed
extends beyond the Galactic disk. White dwarf stars, on the other
hand, are some five magnitudes fainter, and their spatial
proximity allows their numbers to increase steadily with
increasing magnitude until they outnumber the hot subdwarfs by 100
to 1 for $g > 19$. \citet{2004ApJ...607..426K} presented a catalog
of spectroscopically identified WDs in the DR1. In our latest
detection survey of UV-bright stars in the SDSS, we detected a
total of 639 hot subdwarfs and 11\,752 white dwarf stars.

The SDSS spectroscopic survey targets a range of objects based on its properties measured in the photometric {\em ugriz} survey, i.e. depending on whether a source is extended or point-like and its position in color-color space. The survey was optimized for detecting quasars and faint galaxies, but since
these are scattered over significant parts of color-color space, the main strategy of the SDSS selection procedure is to avoid the main stellar loci. It is also worth noting that this procedure became more effective after the early data release (EDR) of the survey, so the selection biases are not constant even within the original SDSS dataset.

Our selection procedure resembles that of the SDSS' own in that we
investigated a large fraction of color-color space in the EDR
release, and refined our search area for the data sets of
subsequent releases. Of particular importance for the WD-M sample
was the cutoff in $g-r$. For the EDR, we imposed no cutoff, but in
later releases we imposed $g-r < 0.2$ in addition to the $u-g <
0.8$ criterion used to select stars with a UV excess. While $g-r <
0.2$ includes effectively all hot subdwarf stars, including
systems with F- to G-class companions, this limit cuts straight
through the locus of WD-dM stars as illustrated in Fig.
\ref{fig:color}. The truncated region of color-color space is also
where the data for the most significant part of the quasars is
found, and the SDSS survey therefore imposes a positive bias on WD
binaries located in this region, while giving lower priority to
objects with negative $g-r$, where single WD stars are found.
However, due to our selection procedure, only a fraction of the WD
binaries with $g-r < 0.2$ is included in this study. Also, the
cool end of the WD population is found to be located redward of
the corner of our selection region at $u-g = 0.8$ and $g-r = 0.2$.
At this point, the WD sequence reaches into the population of
normal halo stars, which the SDSS selection procedure -- as well
as our color-color cuts -- avoids. A more detailed explanation of
the color-color loci of WD-M star systems within the SDSS is given
in \citet{2004ApJ...615L.141S}.

Our sample is made up of 857 objects, which represents our master
sample (see Fig. \ref{fig:color}). We then discard noisy spectra
and those in which either the WD or the red companion have rather
weak features. Only spectra showing either a WD with a
hydrogen-dominated atmosphere (DA white dwarfs) or a WD with clear
helium features (DO or DB white dwarfs) are retained. Furthermore,
we reject spectra with pollution from nearby light sources, which
we verified with the aid of the photometric data. WD-M pairs with
cool WDs and dominant red components are discarded preferentially
from our sample by the selection method, in addition to systems
with significant light contribution due to mass overflow. The
identification of our subsample of 636 systems is therefore
subject to severe selection effects.

\subsection{Premises}
\label{sub:premises}

As a basis for our investigations, we rely on three premises:

\begin{enumerate}[1.]

\item
Due to our model grid resolution of 0.5\,dex for the surface gravity of the M star ($g_{\mathrm{M}}$) and
the low resolution of the SDSS spectra, the uncertainties in our fits for $g_{\mathrm{M}}$ are too large to provide useful constraints on the secondary masses. We assume instead a mass-radius relation for
unevolved main sequence stars by using the evolutionary models of \citet{1997A&A...327.1039C} for a fixed
M star age of $10^{10}$\,yr to deduce the radius ($R_\mathrm{M}$) and mass ($M_\mathrm{M}$) of the M star from its fitted effective temperature ($T_{\mathrm{eff}}^{\mathrm{M}}$) and metallicity [Fe/H]$_\mathrm{M}$. For the mass range considered here, evolution effects on the main sequence are negligible within a Hubble time, thus only pre-main sequence stars with ages $\lesssim 5 \cdot 10^{8}$\,yr might introduce larger errors in the radius estimate. Such young ages, however, can be excluded due to the cooling ages of the white dwarfs in almost all cases.

\item
Similarly, the errors in the surface gravity of the WD
($g_{\mathrm{WD}}$) from our fits are in most cases inadequately
high for a meaningful deduction of the WD radii. However, as shown
by \citet{2007A&A...466..627H}, most of the SDSS DAs have masses
that cluster closely around the peak of the field white dwarf mass
distribution at 0.58\,$M_{\sun}$. There are observational
indications of higher WD masses in magnetic CVs (mCVs) compared with
the masses of field WDs \citep[and references
therein]{2000MNRAS.314..403R, 1998A&AS..129...83R,
1998MNRAS.293..222C}. The maximum of the WD mass function in those
binaries, although not as distinctive as the peak in field-WD
masses, appears to be located between 0.7\,$M_{\sun}$ and
0.8\,$M_{\sun}$. Although we cannot assess the orbital separations of the duets in our sample, except for some known close binaries (PCEBs) and optically resolved duets, we find no indications, such as accretion disk features, for a single mCV in our sample (see Sect. \ref{sec:discussion}). We therefore use a fixed WD mass ($M_{\mathrm{WD}}$) of 0.6\,$M_{\sun}$ and the WD temperature ($T_{\mathrm{eff}}^{\mathrm{WD}}$) derived from the fit to estimate the radius of the WD ($R_\mathrm{WD}$) from the evolutionary tracks calculated by \citet{1994esa..conf..612W}.
These models are available for 0.4, 0.5, 0.6, 0.7, 0.8, and 1.0\,$M_{\sun}$.

\item
The probability of an alignment of a WD and an M star by
chance is found to be negligible. Our reasoning is based on the
fact that our binaries found their way into the sample due to the
high flux contribution of the WD to the blue part of the spectrum.
Hence, WDs are a preferred byproduct of the SDSS and some have M
star companions. We compute the probability $P_\mathrm{M}$ of
finding at least one M star within a circle of diameter equal to
that of an SDSS fiber on the celestial plane, which is
$3\,\arcsec$ \citep{2008ApJS..175..297A}, by chance. It is given
by the Poisson probability of finding $\nu$ objects

  \begin{equation}
  P_\mathrm{M}(\nu) \ = \ \frac{\mu^{\nu}}{\nu!}\,e^{-\mu}
  \end{equation}

and, in our case,

  \begin{equation}
  P_\mathrm{M}(\nu \geq 1) \ = \ 1 - e^{-\mu} \hspace{0.5cm} , \hspace{1cm} \mu \ = \ A \cdot \rho_\mathrm{M}
  \end{equation}

where $A = \pi \cdot (1.5\,\arcsec)^2$ is the probed area and $\rho_\mathrm{M}$ is the area density of M stars on the celestial plane. To estimate $\rho_\mathrm{M}$, we refer to private communication with J. Bochanski, who measured the field luminosity function of stars in the DR6. From there, we derived star
counts in the SDSS \textit{u-, g-, r-, i-} and \textit{z-} filters for the magnitude range of our sample: 15 $<$ \textit{u, g, r, i, z} $<$ 20.5. These counts yield $\rho_\mathrm{M} = 9.55 \cdot 10^{4}\,(\arcsec)^{-2}$, and we get $P_\mathrm{M} = 6.73 \cdot 10^{-3}$ and an expectation value of $P_\mathrm{M} \cdot 636 \approx 4.3$ to be the number of WD-M binaries aligned by chance within our sample. This is why we may use physical interaction of the binary constituents as a basic principle.

\end{enumerate}

\section{Models}
\label{sec:Models}

The central element of our analysis is the synthesis of binary spectra from of theoretical spectra based on model atmosphere calculations for white dwarfs and main sequence stars, respectively. Models for the full parameter range of interest were pre-computed, providing two grids for the spaces of white dwarf
parameters $T_{\mathrm{eff}}^{\mathrm{WD}}$ and $g_{\mathrm{WD}}$ and M dwarf parameters $T_{\mathrm{eff}}^{\mathrm{M}}$, $g_{\mathrm{M}}$ and [Fe/H]$_\mathrm{M}$, respectively. White dwarf spectra were calculated for pure hydrogen atmospheres covering surface gravities of $7 \leq \log(g_{\mathrm{WD}}) \leq 9$ with a step size of 0.5\,dex in a temperature range from 6 to 90\,kK. We adopted steps of 1\,kK between 6 and 30\,kK, steps of 2\,kK between 30 and 50\,kK, and steps of 5\,kK for temperatures above 50\,kK. This core model grid was supplemented with existing models for extremely hot ($T_{\mathrm{eff}}^{\mathrm{WD}} > 90\,\mathrm{kK}$) or lower-gravity ($\log(g_{\mathrm{WD}}) \in \{5,\,6\}$) atmospheres and for hot helium-rich atmospheres. The latter were available for $\log(g_{\mathrm{WD}}) = 7.5$ in the range of $40\,\mathrm{kK} \leq T_{\mathrm{eff}}^{\mathrm{WD}} \leq 80\,\mathrm{kK}$ every 5\,kK. The grid of the MS spectra was almost complete for $2\,600\,\mathrm{K} \leq
T_{\mathrm{eff}}^{\mathrm{M}}$ every 200\,K, [Fe/H]$_\mathrm{M} \in \{-1.5,\,-1,\,-0.5,\,0,\,0.3\}$ and $\log(g_{\mathrm{WD}}) \in \{2,\,3,\,4,\,4.5,\,5,\,5.5\}$.

The white dwarf models were computed by D.~Koester using his codes developed for static, plane-parallel, stellar atmospheres in radiative, hydrostatic, and local thermodynamic equilibrium (LTE), as described in detail in \citet{1997ApJ...488..375F}. The main body of this grid covers fully-blanketed pure hydrogen (DA)
atmospheres. These models consider convective flux according to the mixing length approximation (MLT), by using a variation of the standard formulation of \citet{mihalas78} designated as ML2 in the notation of \citet[and references therein]{1997ApJ...488..375F}, and adopting a mixing length of $\alpha = 0.6$ (in units of pressure scale height). This setup was demonstrated to be one of the most effective available configurations for reproducing DA spectra at high resolution and $S/N$ \citep{1998A&A...338..563H,2001A&A...378..556K}. This grid was extended by an existing set of hydrogen atmosphere models based on the same input physics to cover the more extreme parts of the white dwarf parameter space in high $T_{\mathrm{eff}}^{\mathrm{WD}}$ and lower gravity (sdB-like atmospheres). In addition, for those hot white dwarfs that could not be reproduced accurately by DA spectra the sequence of helium-rich (DO) models was used. These models assume a helium-to-hydrogen mixing ratio of 100:1 and the ML1 MLT version with $\alpha = 1.0$ as described in \citet{1997A&A...318..461J}. These latter models only cover the spectrum up to $\lambda \leq 8\,000\,\AA$, and thus an important part of the dM flux had to be masked out in the analysis. However, none of the primaries were identified to be in the low-gravity or ultra-hot domain of the DAs/sdBs, and a more quantitative analysis of the helium-rich white dwarfs was beyond the scope of this work, as detailed below.

The secondary spectra were calculated with version 14.2 of the multi-purpose stellar atmosphere code \texttt{PHOENIX} \citep{jcam} for 1D spherically symmetric, static atmospheres in LTE, which consider convective instability in the framework of MLT according to \citet{mihalas78} with a mixing length parameter $\alpha = 2.0$. Our models follow the general setup used for the first GAIA grid \citep{inesGAIA}, which included a number of updates from the NextGen grid \citep{1999ApJ...525..871H} and the microphysics described by \citet{2001ApJ...556..357A}, but ignores the effects of condensate formation. We therefore restricted the range of validity of these models to $T_{\mathrm{eff}}\ge 2800$\,K. Major modifications relative to the models of \citet{inesGAIA} included updating of the (scaled) elemental
abundances to the revised solar composition of \citet{agsSolAbun} and considering microturbulent line broadening with a statistical velocity $\chi = 1.0\,\mathrm{km\,s}^{-1}$, which provided a more reliable description of the observed line profiles in cooler stars \citep{jacobMspectra,2008arXiv0805.4826S}.

\section{Mathematical Treatment}
\label{sec:Math}

\begin{figure*}
 \centering
 \scalebox{0.39}{\includegraphics{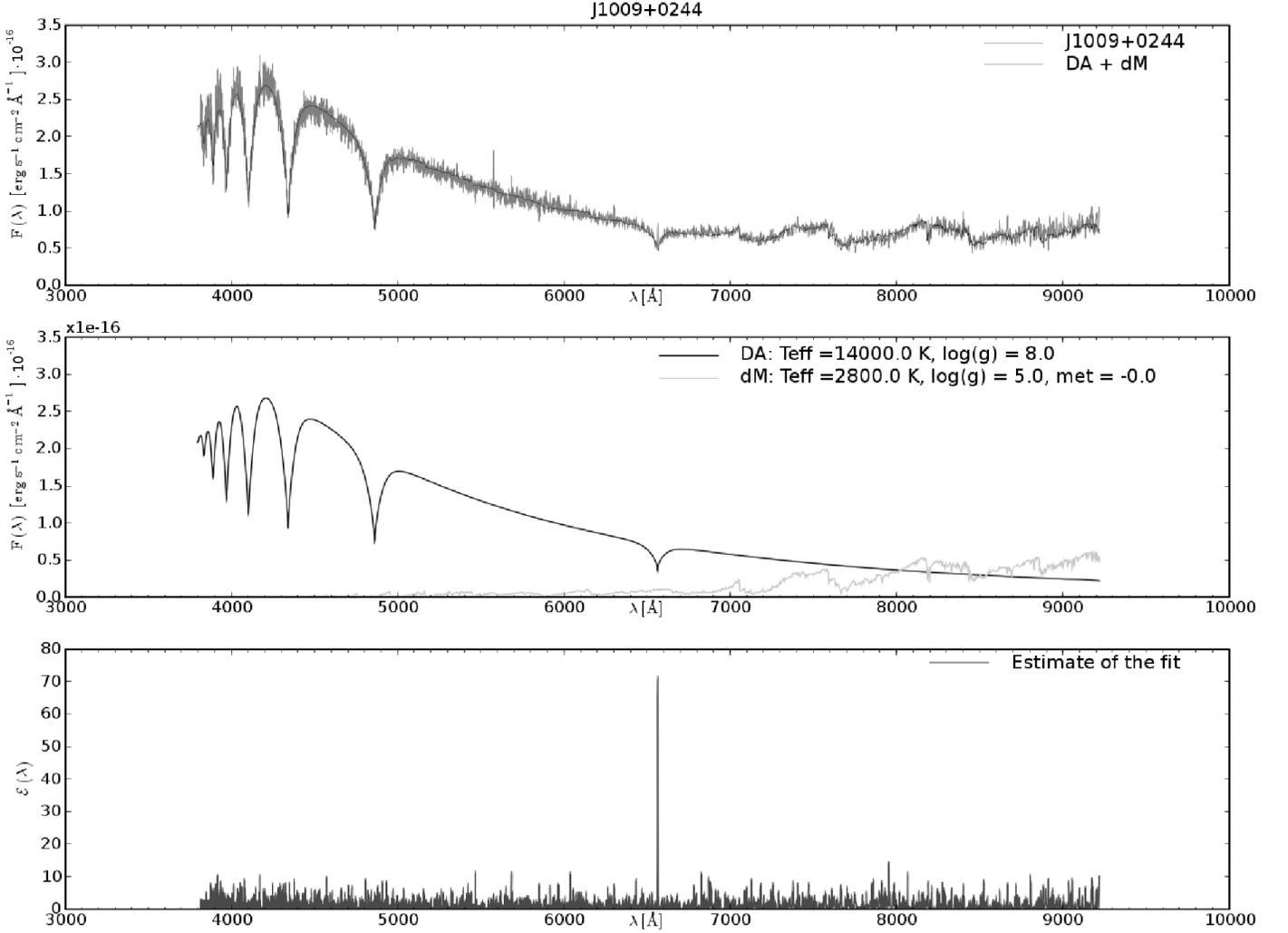}}
 \caption{Example for our fitting procedure; \textit{Top}: Observed SDSS spectrum of J1009+0244 and the DA-dM combination in our fitting routine. As mentioned in \citet{2006AJ....131.1674S}, the feature close to 5\,600\,$\AA$ is an artifact of the SDSS data reductions, and not a stellar feature. \textit{Center}: Decomposition of the observed spectrum. \textit{Bottom}: Estimate of the fit as defined in Eq. (\ref{equ:Estimate}). Note the mismatch of the emission in the H$_\alpha$ line !}
 \label{fig:02}
\end{figure*}

To decompose a composite spectrum, one usually fits a WD model to
its hot region, i.e. the relatively blue wavelength range of the
spectrum, where contamination of the companion star's light is
rather weak. In the next step, this best-fit model for the WD is
subtracted from the combined spectrum to obtain a spectrum most
closely resembling the light of the secondary. Depending on the
density of the model grid and the success of the preceding fit,
one improves the quality of the replicated spectrum with each of
those iteration steps, while $\chi^2$ is reduced, approaching the
local $\chi^2$ minimum in the respective parameter space. Although
this proceeds rapidly, there is the danger of being trapped inside
a local minimum instead of the global minimum. This procedure also
implies a mutual dependence of the computed flux scaling factors.
We therefore propose an alternative technique.

With our $\chi^2$ fitting method, we operate in a five-dimensional parameter space, spanned by $T_{\mathrm{eff}}^{\mathrm{WD}}$, $T_{\mathrm{eff}}^{\mathrm{M}}$, $g_{\mathrm{WD}}$, $g_{\mathrm{M}}$, and [Fe/H]$_\mathrm{M}$. Each observed flux data point $F_{i}^{\mathrm{obs}}$ in a binary spectrum with a total number of $m$ observed data points is reproduced by a combination of two single-star models $x_{i}$ and $y_{i}$. These are weighted with scaling factors $a$ and $b$, respectively, depending on the distances of the stars. With the definition of $\chi^2$ from \citet[Eq.~15.1.5]{NumRec}, we have

\begin{equation}
\label{equ:DefChiSq} \chi^2 \ \coloneqq \ \sum_i^{m}
\frac{(F_{i}^{\mathrm{obs}} -
F_{i}^{\mathrm{mod}})^2}{\sigma_{i}^2} \ = \ \sum_i^{m}
\frac{(F_{i}^{\mathrm{obs}} - a\,x_{i} - b\,y_{i}
)^2}{\sigma_{i}^2}
\end{equation}

\noindent
with $\sigma_{i}$ as the observational error given in the SDSS \textsf{.fits} file of the object. In the following, we use $[w_{i}]$ as an abbreviation for $\sum_i^m w_{i} / \sigma_{i}^2$. We rewrite Eq. (\ref{equ:DefChiSq}) as

\begin{equation}
\chi^2 \ \coloneqq \ [(F_{i}^{\mathrm{obs}} - a\,x_{i} -
b\,y_{i})^2] \hspace{0.5cm} . \tag{3}
\end{equation}

\noindent Since we attempt to identify that model combination of
models $x$ and $y$ that corresponds to the global $\chi^2$
minimum, we have the boundary conditions

\begin{align}
0 \ & \stackrel{!}{=} \ \nabla\,\chi^2(a,b) \ = \ \left(
\frac{\partial}{\partial a}\,\chi^2(a,b) \ , \
\frac{\partial}{\partial b}\,\chi^2(a,b) \right) \hspace{0.5cm}
\mathrm{and}\\ \nonumber 0 \ & < \ \Delta\,\chi^2(a,b)
\hspace{0.5cm} .
\end{align}

\noindent
With

\begin{equation}
\nabla \ \chi^2(a,b) \ = \ -2\,\left([x_{i}\,(F_{i}^{\mathrm{obs}}
- a\,x_{i} - b\,y_{i})] \ , \ [y_{i}\,(F_{i}^{\mathrm{obs}} -
a\,x_{i} - b\,y_{i})] \right)
\end{equation}\vspace{0cm}

\noindent
we obtain

\begin{equation}
  \begin{pmatrix}
    [F_{i}^{\mathrm{obs}}\,x_{i}]\\
    [F_{i}^{\mathrm{obs}}\,y_{i}]\\
  \end{pmatrix} \
= \ \underbrace{
  \begin{pmatrix}
    [x_{i}^2]\,[x_{i}\,y_{i}]\\
    [y_{i}\,x_{i}]\,[y_{i}^2]\\
  \end{pmatrix}
}_{Z}
  \begin{pmatrix}
    a\\
    b\\
  \end{pmatrix} \hspace{0.5cm} ,
\end{equation}

\noindent
which is equivalent to

\begin{equation}
  \begin{pmatrix}
     a\\
     b\\
   \end{pmatrix} \
= \ Z^{-1}
  \begin{pmatrix}
    [F_{i}^{\mathrm{obs}}\,x_{i}]\\
    [F_{i}^{\mathrm{obs}}\,y_{i}]\\
  \end{pmatrix}
\hspace{0.5cm} .
\label{equ:scale}
\end{equation}

\noindent
This finally leads us to

\begin{equation} \label{equ:ScaleSolution}
  \begin{pmatrix}
    a\\
    b\\
  \end{pmatrix} \
= \
  \begin{pmatrix}
    [F_{i}^{\mathrm{obs}}\,x_{i}]\,[y_{i}^2] - [x_{i}\,y_{i}]\,[F_{i}^{\mathrm{obs}}\,y_{i}]\\
    [F_{i}^{\mathrm{obs}}\,y_{i}]\,[x_{i}^2] - [y_{i}\,x_{i}]\,[F_{i}^{\mathrm{obs}}\,x_{i}]\\
  \end{pmatrix}
\frac{1}{[x_{i}^2]\,[y_{i}^2] - [x_{i}\,y_{i}]^2} \hspace{0.5cm} .
\end{equation}\vspace{0cm}

\noindent Equation (\ref{equ:ScaleSolution}) provides the scaling
factors for both the WD and the M star model, which we then use to
compute $\chi^2$ with Eq. (\ref{equ:DefChiSq}). Using this
procedure, we are able to avoid a mutual dependence of $a$ and $b$
since the system of equations can be solved uniquely, and we also
avoid the effects of identifying a local, and not the global,
$\chi^2$ minimum. However, the technique is more time consuming
since we must first compute the $\chi^2$ distribution for the
complete parameter ranges that we wish to consider.

\begin{figure}
  \centering
  \scalebox{0.37}{\includegraphics{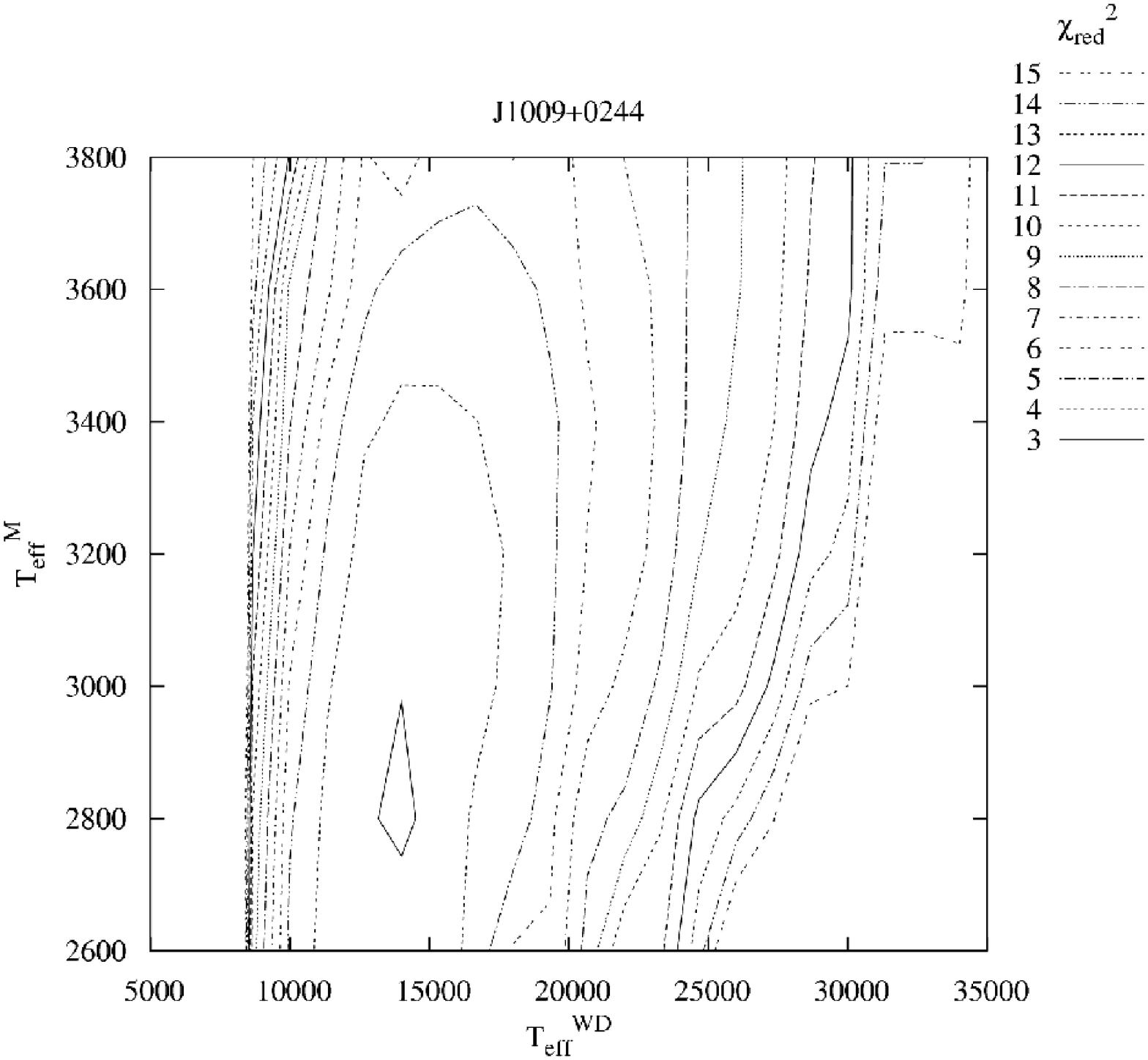}}
  \caption{Projection of the $\chi^2$ landscape onto the $T_{\mathrm{eff}}^{\mathrm{WD}}$-$T_{\mathrm{eff}}^{\mathrm{M}}$ plane for J1009+0244; The best fit in this case was found at $T_{\mathrm{eff}}^{\mathrm{WD}} = 14\,000$\,K and $T_{\mathrm{eff}}^{\mathrm{M}} = 2\,800$\,K.}
 \label{fig:J1009+0244_TWD-vs-TM_contour}
\end{figure}

\noindent
Additionally, we introduce $\mathcal{E}$ as an estimate of the fit, which is defined to be

\begin{equation}\label{equ:Estimate}
\mathcal{E} (\lambda) \ \coloneqq \ \frac{ \left( F_{\mathrm{obs}} (\lambda) - F_{\mathrm{mod}} (\lambda) \right)^2 }{\sigma (\lambda)^2}
\end{equation}

\noindent and resembles the definition of $\chi^2$ in Eq.
(\ref{equ:DefChiSq}). Since we do not evaluate the sum over all
data points $i$, the estimate $\mathcal{E} (\lambda)$ may be
regarded as $\chi^2_{\mathrm{red}} (\lambda)$ or, in other words,
as an individual estimate of each single data point $i$. It
provides a means of applying a manual cutoff-mask to noisy parts
of the spectrum, implemented in our fitting routine. Hence, we
achieve the decomposition of the combined spectrum, and an example
of a typical binary spectrum in our sample can be seen in
Fig.~\ref{fig:02}. In Fig.~\ref{fig:J1009+0244_TWD-vs-TM_contour},
we show an example of the projection of the five-dimensional
$\chi^2$ landscape onto the
$T_{\mathrm{eff}}^{\mathrm{WD}}$-$T_{\mathrm{eff}}^{\mathrm{M}}$
plane. We use

\begin{align}
\sigma_{i,j} \ & = \ \sqrt{ \frac{d_{i,j}^2}{\chi_{j}^2 - \chi^2_{\mathrm{min}}} } \hspace{0.5cm} ,\\ \nonumber
i & \in \{T_{\mathrm{eff}}^{\mathrm{WD}}, \ T_{\mathrm{eff}}^{\mathrm{M}}, \ g_{\mathrm{WD}}, \ g_{\mathrm{M}}, \ \mathrm{[Fe/H]}_{\mathrm{M}}  \}, \ j \in \{\mathrm{up, \ down}\}
\label{equ:Zhang_1986}
\end{align}

\noindent
given in \citet{1986ApJ...305..740Z} to calculate the 1\,$\sigma$-confidence intervals of $\approx 68$\,\% for our fits. The lower and upper grid step widths in the $i$th parameter dimension about the $\chi^2$ minimum are given by $d_{i,j}$. For example,
$d_{T_{\mathrm{eff}}^{\mathrm{M}}, \mathrm{down}} \equiv 200 \,
\mathrm{K} \equiv d_{T_{\mathrm{eff}}^{\mathrm{M}}, \mathrm{up}}$.

We define $\mathfrak{p}$ to be the number of parameters, such that
we have $\mathfrak{p} = 5$, and $m$ as the number of data points
in an SDSS spectrum. We have mainly $m \approx 3830$, but, in some
cases, parts of a spectrum must be discarded, then $m$ becomes
smaller. We denote the number of statistically independent data
points in the spectrum for an inverse resolution $R$, by

\begin{equation}
N \ = \ \int_{N_\mathrm{min}}^{N_\mathrm{max}} \mathrm{d} N \ = \ R \
\mathrm{ln} \left(
\frac{\lambda_\mathrm{max}}{\lambda_\mathrm{min}} \right)
\hspace{0.5cm} .
\end{equation}

\noindent where $\lambda_\mathrm{min/max}$ are the lower and upper
wavelength limits of 3\,800\,{\AA} and 9\,200\,{\AA}, yielding
between 1\,600 and 2\,000 resolution elements per spectrum. We
define $n$ to be the number of degrees of freedom given by $n = N
- \mathfrak{p}$, which is the number required for our mathematical
evaluations. To ensure that the quality of the fits are comparable
with each other, we compute the reduced $\chi^2$ via

\begin{equation}\label{equ:ChiSqRed}
\chi^2_{\mathrm{red}} \ \coloneqq \ \frac{\chi^2}{n}
\hspace{0.5cm} .
\end{equation}

\noindent
We use the concept of the quantile that provides an
estimation of how probable a computed $\chi^2_\mathrm{min}$ is
depending on the number of degrees of freedom $n$, and is defined
to be

\begin{equation}\label{equ:quantile}
p_{n}(\chi^2 \le \chi^2_\mathrm{min}) \ = \
\frac{1}{2^{n/2}\,\Gamma(\frac{n}{2})}
\int_{0}^{\chi^2_\mathrm{min}} \mathrm{d} \chi^2 \ (\chi^2)^{n/2 -
1}\,e^{- \chi^2 / 2} \hspace{0.5cm} ,
\end{equation}

\noindent
which provides the probability of $\chi^2$ being equal or smaller than a certain $\chi^2_\mathrm{min}$.

\subsection{Distances}

When $R_{\mathrm{WD}}$ is inferred from the fixed value of $M_{\mathrm{WD}}$, the fitted
$T_{\mathrm{eff}}^{\mathrm{WD}}$, and by using models from \citet{1994esa..conf..612W}, we derive the distance $d_{\mathrm{WD}}$ of the WD from Earth by obtaining the scaling factor $a$ from our fitting method, which scales the model flux to the observed flux. Since $a = (R_\mathrm{WD} / d_\mathrm{WD})^2$, we have

\begin{equation}
d_{\mathrm{WD}}\,[\mathrm{pc}] \ = \ \frac{
R_{\mathrm{WD}}\,[R_{\sun}]}{1\,\mathrm{pc} \, \sqrt{a} / R_{\sun}}
\hspace{0.5cm} .
\end{equation}

\begin{figure}
 \centering
 \scalebox{0.17}{\includegraphics{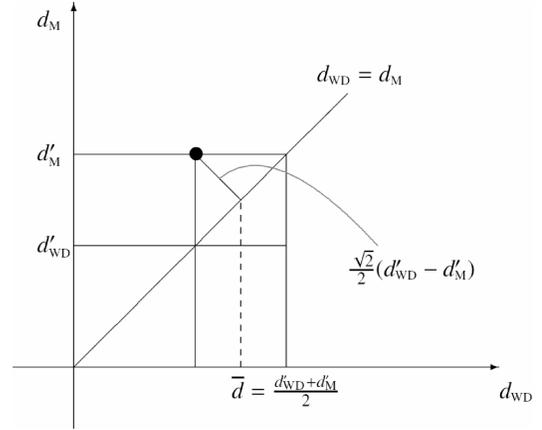}}
 \caption{Geometrical explanation of the distance evaluation coefficient $\mathcal{C}$.}
 \label{fig:dist_expl}
\end{figure}

\noindent
A similar procedure is applied to deduce the distance $d_\mathrm{M}$ of the M star. These two distances should be equal ideally in case the two stars form a physical pair. In a plot with $d_\mathrm{WD}$ on the abscissa and $d_\mathrm{M}$ on the ordinate, physical systems should lie on or near the diagonal where $d_\mathrm{WD} = d_\mathrm{M}$. Their displacement from the diagonal is given by $\sqrt{2}\,(d_\mathrm{WD} - d_\mathrm{M}) / 2$. To ensure that the estimates of the respective distances are comparable for all the spectra, we introduce the coefficient $\mathcal{C}$. It weighs the displacement of a certain binary system from the diagonal by the average distance of the system from Earth, which is $\overline{d} = (d_\mathrm{WD} + d_\mathrm{M}) / 2$ (see Fig. \ref{fig:dist_expl}):

\begin{equation}
\mathcal{C} \coloneqq \frac{  \sqrt{2}\,(d_\mathrm{WD} -
d_\mathrm{M}) / 2 }{ (d_\mathrm{WD} + d_\mathrm{M}) / 2 } \ = \
\sqrt{2}\,\frac{d_\mathrm{WD} - d_\mathrm{M}}{d_\mathrm{WD} +
d_\mathrm{M}}
\label{equ:C}
\end{equation}

\noindent
Thus, a discrepancy between $d_\mathrm{WD}$ and $d_\mathrm{M}$ is more significant when the binary is closer, and we expect $\mathcal{C}$ to be close to zero for those cases in which our premises are justified and the spectral fit is good. The algebraic sign indicates the direction of the displacement.

\subsection{Optically Resolved Binaries}

In addition to the spectroscopic data, we also studied the
photometric data. Since we hardly detect the smeared Airy disks of
the objects, we are unable to derive any morphological
information, but a possible separation of the spot centers allows
us to estimate their projected mutual distance
$d_{\mathrm{proj}}$. For this purpose, we use the averaged
distance $\overline{d}$ of the system from Earth, the displacement
$D$ of the stars on the SDSS images in units of pixels, and the
given resolution $\varrho$ of the images in $\arcsec/$pixel. We
calculate the angular separation $\alpha$ in units of radians by
means of $\alpha = \varrho\,D / 3600 \cdot \pi / 180$ and by using
geometry,

\begin{equation}
d_{\mathrm{proj}} \ = \ 2\,\overline{d}\,\tan \left(
\frac{\alpha}{2} \right) \hspace{0.5cm} .
\end{equation}

\begin{figure}
 \centering
 \scalebox{0.14}{\includegraphics{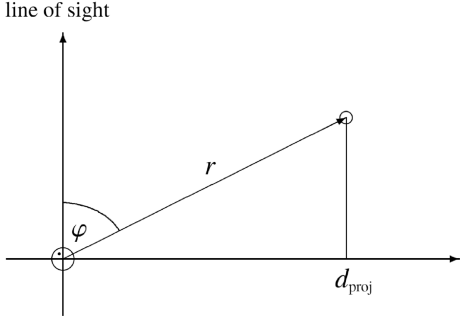}}
 \caption[Projection of the displacement $r$ of two stars]{$d_{\mathrm{proj}}$ is the observed projection of the displacement $r$ of the two stars. Averaged over many observations, the latter is about 1.57 times smaller than the real spatial separation.}
 \label{fig:projection}
\end{figure}

\noindent
The projected distance, however, is shorter than the true spatial separation $r$ (see Fig. \ref{fig:projection}). If we knew $r$ of a system and were able to measure $d_{\mathrm{proj}}$ many times during an orbit, or if we knew $r$ for many systems and were able to measure the respective $d_{\mathrm{proj}}$ for each system, then we would state that, on average, $d_{\mathrm{proj}}$ reaches a value closer to $0.637\,r$, the more observations that we have. We can deduce this mathematically by considering the mean value of $d_{\mathrm{proj}}$, which is

\begin{equation}
<d_{\mathrm{proj}}> \ = \ <r\,\sin(\varphi)> \ = \
r\,<\sin(\varphi)> \hspace{0.5cm} ,
\end{equation}

\noindent
where the mean of the sine of the projection angular $\varphi$ is given by

\begin{equation}
<\sin(\varphi)> \ = \ \frac{\displaystyle{\int_0}^\pi \mathrm{d}
\varphi\,\sin(\varphi)}{\displaystyle{\int_0}^\pi \mathrm{d}\varphi} \ = \
\frac{2}{\pi} \ \approx \ 0.637 \hspace{0.5cm} .
\end{equation}

\noindent
This implies, that the true spatial separation is, on average, about $1/0.637 \approx 1.571$ times longer than the observed projected distance. We will denote this statistical correction of $d_{\mathrm{proj}}$ with
$d_{\mathrm{proj}}^{\mathrm{corr}} \coloneqq 1.571\,d_{\mathrm{proj}}$.

Using the optical separation of the stars, we can use the visible separation and estimate the period of the system by means of

\begin{equation}
P \ \gtrsim \ 2 \pi\,\sqrt{ \frac{d_{\mathrm{proj}}^3}{ G\,( M_{\mathrm{WD}} + M_{\mathrm{M}} ) } } \hspace{0.5cm} , \label{equ:period}
\end{equation}

\noindent
where we assume that $M_\mathrm{WD} = 0.6\,M_{\sun}$ on the basis of arguments given above and for the sake of consistency with the fitting procedure. We derive $M_{\mathrm{M}}$ from evolutionary tables calculated by \citet{1998A&A...337..403B} (see Sect. \ref{sub:premises}). Equation (\ref{equ:period}) may only provide an approximate approach, due not only to the uncertainties in the masses but also because of the approximation of the projection of the mutual distance is smaller and almost equal to the sum of the major
axes $a_\mathrm{WD}$ and $a_\mathrm{M}$ of the ellipses on which the stars move: $d_{\mathrm{proj}} \lesssim a_\mathrm{WD} + a_\mathrm{M}$. If we take into account the averaged correction of the projection, then the corrected period $P^{\mathrm{corr}}$ is longer than $P$. We note that deviations $\mathrm{d} M_\mathrm{WD}$ from $M_\mathrm{WD} = 0.6\,M_{\sun}$ have a significant effect, since the mass was already used for the derivation of $d_{\mathrm{proj}}$. Since $d_{\mathrm{proj}}^3 \propto r_\mathrm{WD}^3 \propto M_\mathrm{WD}^{-1}$, we derive

\begin{equation}
\frac{\mathrm{d} P}{\mathrm{d} M_\mathrm{WD}} \ \propto \ \frac{\mathrm{d}}{\mathrm{d} M_\mathrm{WD}} \left( \frac{M_\mathrm{WD}^{-1}}{M_\mathrm{WD} + M_\mathrm{M}} \right)^{1/2} \hspace{0.5cm} ,
\end{equation}

\noindent
which leads to

\begin{align}
\ \mathrm{d} P \ & \propto \ - \frac{ 2\,M_\mathrm{WD} + M_\mathrm{M}}{ 2\,\left[ M_\mathrm{WD}\,(M_\mathrm{WD} + M_\mathrm{M})\right]^{3/2}} \, \mathrm{d} M_\mathrm{WD}&\\ \nonumber
& \approx - M_\mathrm{WD}^{-2}\,\mathrm{d} M_\mathrm{WD} \hspace{3.6cm} | \ M_\mathrm{WD} > M_\mathrm{M} \hspace{0.5cm} .
\end{align}

\noindent
Thus, deviations in the period are more significant when the mass of the white dwarf is lower and, for $M_\mathrm{WD} = 0.6\,M_{\sun}$, they could be considerable. However, since $\mathrm{d} M_\mathrm{WD}$ is usually small, impacts due to variations in the WD mass should not have a substantial impact on the assessment on the lower limit to $P$, i.e. approximately 10\,\% for $P$.

\section{Results}
\label{sec:results}

The temperature function for our WD sample can be seen in Fig. \ref{fig:Teff_WD}. As mentioned above, it is subject to considerable selection effects. While the strong peak at 17\,000\,K belongs
to the DAs, the second and smaller peak at 40\,000\,K, which is the lowest temperature in our DO model grid, stems from the cool helium-rich WDs. We convolved the original data distribution with a Gaussian probability function about each data point, to account for the uncertainty in the derived parameters. As a measurement of standard deviation, we used 2\,000\,K and normalized the distribution to 1. Since most of the WDs in our sample have a surface gravity of between 7 and 8\,dex (see Fig. \ref{fig:logg_WD}), the shape of the WD-distance function in Fig. \ref{fig:dist_WD} is similar to that of the temperature function.

In Fig. \ref{fig:Teff_M}, we show the temperature function for the M stars. Although the standard
deviation is merely a tenth of that for $T_{\mathrm{eff}}^{\mathrm{WD}}$, the function is not so smooth because of the relatively low number of model grid points. The tail towards temperatures higher than those of M stars is due to 4 objects at $T_{\mathrm{eff}}^{\mathrm{M}} = 4\,000$\,K with insecure fits, i.e. because of He emission in the spectrum or large distances to Earth and therefore weak spectral features. Our results for the other free parameters, $g_{\mathrm{M}}$ and [Fe/H]$_{\mathrm{M}}$, can be seen in Figs. \ref{fig:logg_M} and \ref{fig:Fe-H}, while the distance distribution for the M stars is given in Fig. \ref{fig:dist_M}.

As shown in \citet{2007A&A...466..627H}, the standard deviation from the maximum of the WD mass function at $0.58\,M_{\sun}$ is $\approx 0.1\,M_{\sun}$. This corresponds to an error in our WD distances of $\approx 17\%$, and this uncertainty creates dispersion in the data from the diagonal in the $d_{\mathrm{WD}}$-$
d_{\mathrm{M}}$ plot, namely

\begin{equation}
 \mathcal{C} \ = \ \sqrt{2}\,\frac{0.17 \ d_{\mathrm{WD}}}{d_{\mathrm{WD}} + d_{\mathrm{M}}}
\end{equation}

\noindent
and for physical systems with $d_{\mathrm{WD}} \approx d_{\mathrm{M}}$

\begin{equation}
 \mathcal{C} \ \approx \ \sqrt{2}\,\frac{0.17}{2} \ \approx \ 0.12 \hspace{0.5cm} .
\end{equation}

\begin{figure*}
  \centering
  \scalebox{0.36}{\includegraphics{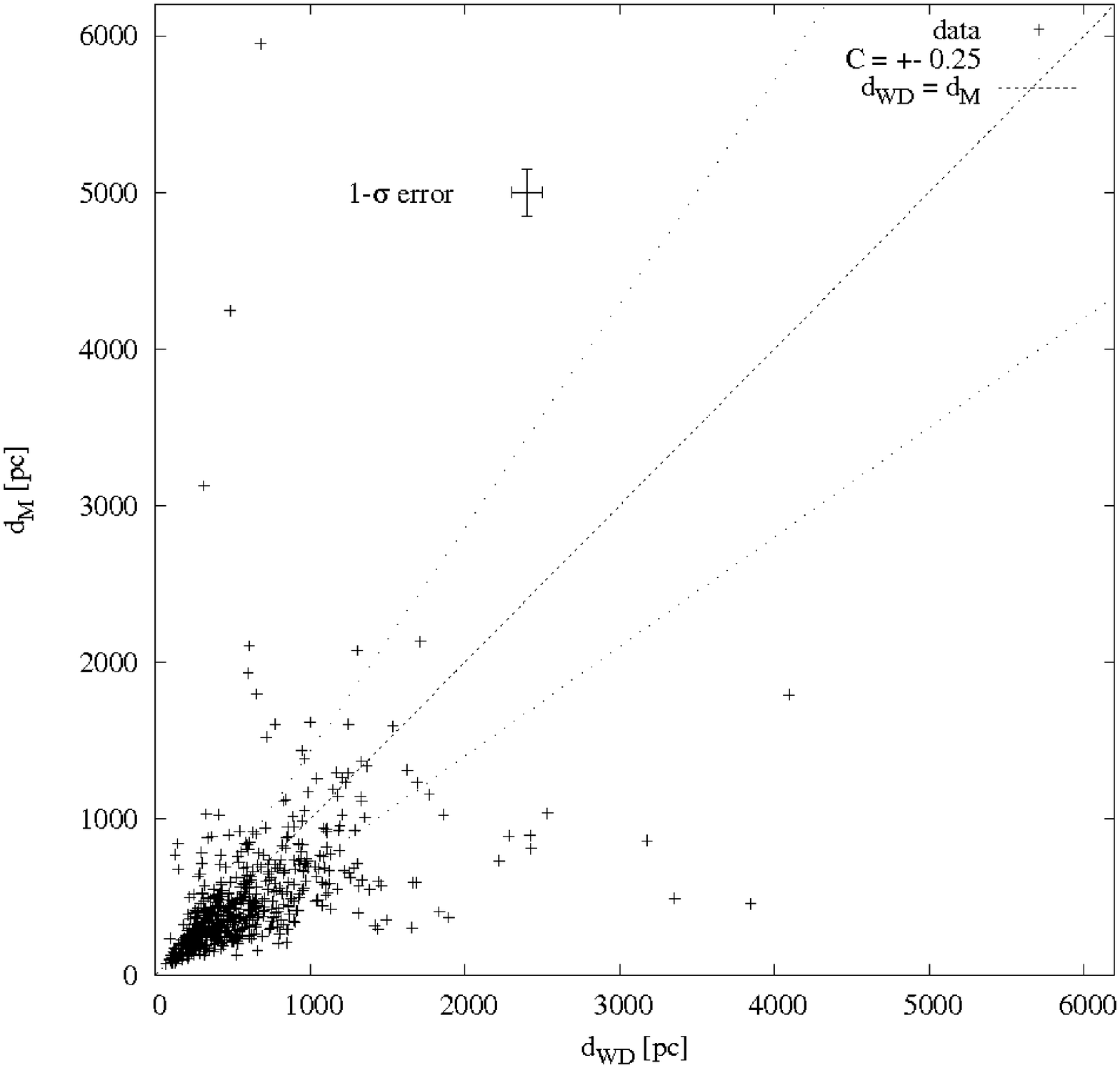}}
  \scalebox{0.36}{\includegraphics{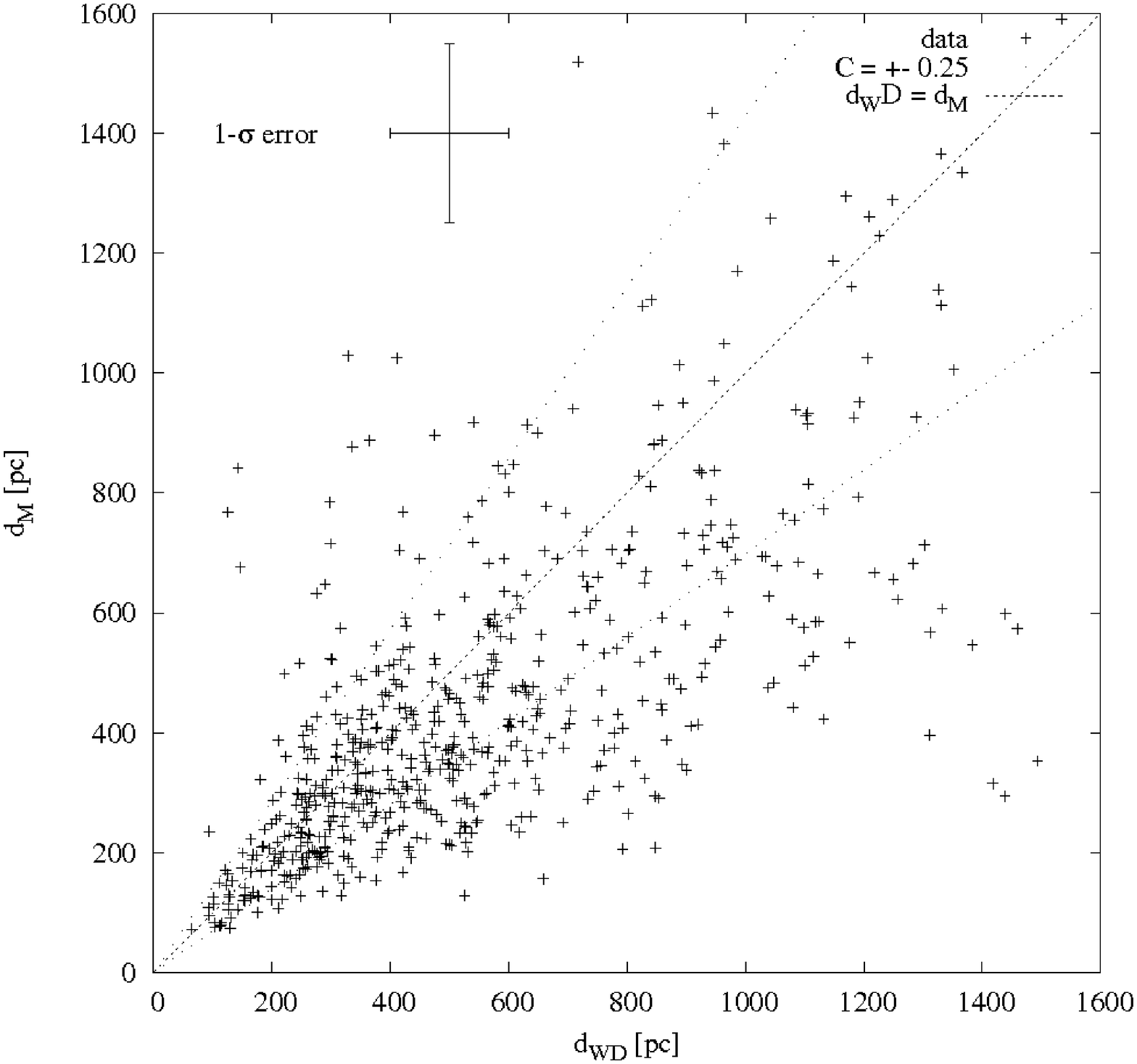}}
   \caption{Distribution of the derived distances of the stellar
components of our systems; The diagonal is the ideal for
physical binaries. The dotted curves span a
tolerance fan for $\mathcal{C} = 0.25$. \textit{Left}: Note that
for very far away -- and thus rather faint -- objects with a
distance of more than $\approx 1\,000$\,pc the distance for the
respective binary mate is typically very different. This is
due to the weak spectroscopic features of the distant stars and
the low spectral resolution. \textit{Right}: Same as left but zoomed in.}
  \label{fig:dist}
\end{figure*}

\begin{figure}
  \centering
  \scalebox{0.35}{\includegraphics{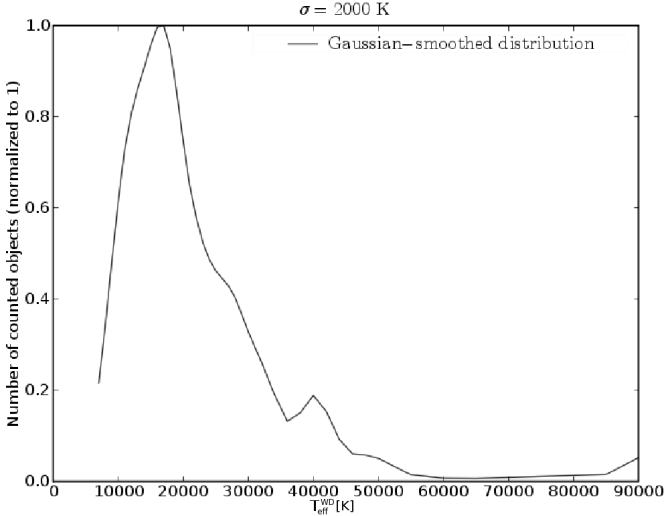}}
  \caption{Temperature function for the WDs; While the major peak at 17\,000\,K belongs to the DAs, the bump at 40\,000\,K is due to the preferential selection of cool DOs for our sample. This plot looks very similar to that shown in \citet{2006AJ....131.1674S}, except for the DO feature and that it is smoothed.}
  \label{fig:Teff_WD}
\end{figure}

\noindent
Taking into account all the other error sources mentioned before, we would expect the systems to be mainly distributed within the fan of $\mathcal{C} = 0 \pm 0.25$ to both sides of the diagonal. In Fig. \ref{fig:dist}, we present the derived distances for all systems. Indeed, this plot, and in particular the zoom-in section in the right panel, shows a scattered distribution, fanned out to both sides of the diagonal, with most of the binaries located in the $\mathcal{C} = 0.25$-tolerance interval. However, we also detect a trend towards higher distances for the WD component -- the scattering is asymmetric with respect to the bisecting line.

For all of the analyzed spectra, $p_n = 1$ (see Eq.
(\ref{equ:quantile})), which implies that a $\chi^2$ equal to or
smaller than the calculated $\chi_{\mathrm{min}}^2$ is probable
for the respective number of degrees of freedom. In other words,
the quality of our fits is poor in mathematical context. The
standard deviations of the measured parameters are quite weak in
terms of physical significance. As an example, we consider
J1009+0244 (see Fig. \ref{fig:J1009+0244_TWD-vs-TM_contour}). We
compute $\sigma_{T_{\mathrm{eff}}^{\mathrm{WD}}, \mathrm{down}}
\approx 80\,\mathrm{K} \approx
\sigma_{T_{\mathrm{eff}}^{\mathrm{WD}}, \mathrm{up}}$, and
$\sigma_{T_{\mathrm{eff}}^{\mathrm{M}}, \mathrm{down}} \approx
5\,\mathrm{K} \approx \sigma_{T_{\mathrm{eff}}^{\mathrm{M}},
\mathrm{up}}$, which demonstrates that the systematic errors are
larger than the mathematical ones. For a conservative estimate of
the errors due to the incomplete molecular data in the M star
models, SDSS flux calibration errors and interstellar reddening,
we refer to \citet{2006A&A...454..617H}, and assume an uncertainty
of 2\,000\,K for the WDs with temperatures smaller than 50\,000\,K
and a 1-$\sigma$ interval of 100\,K for the MS stars. For
$T_{\mathrm{eff}}^{\mathrm{WD}} > 50\,000$\,K, the absolute value
of our accuracy is given by half of the model step width of
5\,000\,K. Our accuracy in the surface gravities is limited by the
low resolution of the SDSS spectra and the step size of our model
grid, and we therefore have $\sigma_{\log(g_{\mathrm{WD}})}
\approx 0.5\,$dex $ \approx \sigma_{\log(g_{\mathrm{M}})}$
\footnote{\citet{2005AN....326..930R} provided an even more
conservative error estimate of 1\,dex, due to the low resolution
of the spectra.}. The metallicity determination is only accurate
to about 0.3\,dex.

\subsection{White Dwarfs Showing He Lines}

From our original input sample, 616 spectra can be fitted well using a DA primary. However, among the remaining systems, He features are evident in a small number in the primary spectrum. These are most readily apparent at relatively high $T_{\mathrm{eff}}^{\mathrm{WD}}$, since spectra with distinctive DB
features were efficiently removed in the initial selection. We found that 20 of the remaining spectra, which were dominated mainly by the He\,\textsc{ii} Bracket-equivalent series, were fitted more accurately by spectra from our DO model library. These stars cluster predominantly at the cool end of the DO sequence: 16 of 20 are fitted with the lowest effective temperature in the DO grid of 40\,000\,K. Many of the them show stronger He\,\textsc{i} lines in addition to He\,\textsc{ii}, and would probably be classified more accurately as DBO, according to the nomenclature of \citet{1993PASP..105..761W}. We suspect that this distribution does not reflect the true luminosity function of Helium-rich white dwarfs, but is rather biased by a contribution of WDs below the $T_{\mathrm{eff}}$ limit of our DO models. Since the number of these systems, and an additional number of potential DB systems that could neither be fitted by DA nor DO models and were omitted from our catalog, is too small for a meaningful statistical analysis, we have not attempted to obtain more reliable fits with additional DB models.

Three DOs in our sample (J0756+4216, J0916+0521, and J1336-0131) were already studied by \citet{2006A&A...454..617H}. For the first one, the parameter values derived in our study are in good agreement with those of the cited publication. While they found the white dwarf to show an effective temperature of $52.5 \pm 0.8$\,kK at $g_{\mathrm{WD}} = 7.4 \pm 0.05$\,dex, we find $T_{\mathrm{eff}}^{\mathrm{WD}} = 50 \pm 5$\,kK and $g_{\mathrm{WD}} = 7.5 \pm 0.5$\,dex. For the MS companion, they derived $T_{\mathrm{eff}}^{\mathrm{M}} = 3.2 \pm 0.2$\,kK, $g_{\mathrm{M}} = 5 \pm 0.5$\,dex, and $M_\mathrm{M} = 0.18\,M_{\sun}$, whereas we measure $T_{\mathrm{eff}}^{\mathrm{M}}
= 3.4 \pm 0.1$\,kK, $g_{\mathrm{M}} = 5 \pm 0.5$\,dex, and $M_\mathrm{M} = 0.27\,M_{\sun}$.

The second binary, J0916+0521, was proposed by \citet{2006A&A...454..617H} to be one of the few optical pairs aligned by chance. With $\mathcal{C} = 0.461$ for that system, we tend to confirm their findings although the data points for many systems are located in the same region of the $d_{\mathrm{WD}}$-$d_{\mathrm{M}}$ diagram around J0916+0521. The other parameter values do not differ
significantly from each other, except for the M star masses, for which they derived $0.51\,M_{\sun}$, while we find $M_{\mathrm{M}} = 0.18\,M_{\sun}$ due to $\approx$\,10\% lower $T_{\mathrm{eff}}$ and subsolar metallicity found in our fit. Our results for the third system, J1336-0131, agree with the values given by \citet{2006A&A...454..617H}.

\subsection{Spectra with Emission}
\label{subsec:emission}

In the spectra of 167 objects in our sample we find at least H$_\alpha$ emission, and some of them also exhibit other types of Balmer emission. Most of these systems probably harbor an active M dwarf with chromospheric emission, or the respective M star might be irradiated by the primary, and experience photoionization and recombination. These systems are non-accreting since they do not show He\,\textsc{i} or He\,\textsc{ii} features. A more detailed treatment of emission in WD-M star binaries is given in \citet{2006AJ....131.1674S}. For an assessment of the magnetic activity, we scanned our fits for pronounced peaks in $\mathcal{E}$, as defined by Eq. (\ref{equ:Estimate}). An example is shown in the lower panel of Fig. \ref{fig:02}. Our M dwarf models did not consider H$_\alpha$ emission, and therefore the spike in the fitting error at the H$_\alpha$ wavelength was -- if present at all -- always related to an emission feature in that line. We decided to attribute the peak in $\mathcal{E}$ close to 6\,563\,$\AA$ to H$_\alpha$ emission, if it exceeded the ambient noise level by at least a fraction of three.

In their study of 101 WDMSs, \citet{2007MNRAS.382.1377R} identified 18 PCEBs and
PCEB candidates of which 12 are included in our master sample (see Table \ref{tab:master_sample}). Eleven of them are in our spectroscopic sample and 8 of them are marked as H emitters therein. Due to the strong $H_{\alpha}$ contamination of the PCEB (candidate) spectra our fits are often inconsistent in terms of $d_\mathrm{WD}$ and $d_\mathrm{M}$, whereas they should ideally be
equal. If the parameter values of our best-fit solution differed
significantly from those provided by \citet{2007MNRAS.382.1377R},
we chose to adopt instead the parameters corresponding to their
solutions.

As an example, we refer to J0052$-$0053, which shows an ambiguous
$\chi^2_{\mathrm{red}}$ distribution. This is probably due to the
weak WD and M star features in the spectrum and due to the large distance of the system. The results of our fitting routine for J0052$-$0053 are $T_{\mathrm{eff}}^{\mathrm{WD}} =
29\,000$\,K, log $g_\mathrm{WD} = 8.5$, and a distance to Earth of
1\,612\,pc for the WD. The corresponding parameters for its
companion are computed to be $T_{\mathrm{eff}}^{\mathrm{M}} =
3\,200$\,K, log $g_\mathrm{M} = 5.0$, [Fe/H]$_\mathrm{M}$ = 0.0,
and $d_{\mathrm{WD}} = 223$\,pc. The respective distance
evaluation coefficient of $\mathcal{C} = 1.07$ emphasizes the
grave dissonance in the objects' distances from Earth. It should
be close to zero for consistent fits as defined in Eq.
(\ref{equ:C}). Hence, we fix the parameters for the WD to be
$M_\mathrm{WD} = 1.0\,M_{\sun}$, $T_{\mathrm{eff}}^{\mathrm{WD}} =
15\,000$\,K and log $g_\mathrm{WD} = 8.5$  -- following
\citet{2007MNRAS.382.1377R} -- and set
$T_{\mathrm{eff}}^{\mathrm{M}} = 3\,600$\,K to enforce
$d_{\mathrm{M}} \approx d_{\mathrm{WD}}$. Using this procedure, we
deduce $d_{\mathrm{WD}} = 579.9$\,pc and $d_{\mathrm{M}} =
577.2$\,pc, in good agreement with the value of 505\,pc for the
distance of the WD derived by \citet{2007MNRAS.382.1377R} (see
Table \ref{tab:results} for the whole result).

We also find 4 systems showing He\,\textsc{i} and He\,\textsc{ii} emission, in addition to H$_\alpha$ features, that harbor a hot primary ($T_{\mathrm{eff}}^{\mathrm{WD}} \gtrsim 65\,000$\,K). Since He\,\textsc{i} is not typically observed in active M dwarfs, these objects might either be CVs, i.e. binaries that undergo mass transfer, or their mutual separation might just be large enough as to avoid mass overflow on the one hand but close enough to produce the emission lines by irradiation effects on the other hand. J1249+0357 was actually classified as CV by \citet{2001PASP..113..764D} and \citet{2004AJ....128.1882S}. However, its CV status can be doubted since the spectrum shows no features of an accretion disk and the He emission lines are not dynamically broadened by a potentially rotating disk. The emission features in J1317+6731 and J1439$-$0106 are most likely due to irradiation of the M component. Their spectra show strong H$_\alpha$ as well as He\,\textsc{i} and He\,\textsc{ii} emission. Both objects were already mentioned in \citet{2006AJ....131.1674S} and the detection of the latter one was originally published in \citet{2003AJ....125.2621R}. The system J2125$-$0107 that we fitted with a DO primary from our model grid is actually known to be a close system with a PG1159 primary \citep{2006A&A...448L..25N, 2008AN....329..376S, 2009Schuh}. We misclassified that object since we did not include PG\,1159 models in our repertoire. Our value for $T_{\mathrm{eff}}^{\mathrm{M}}$ of 4\,000\,K should be taken with reservation due to the known strong irradiation processes from the hot primary on the MS companion.

From the 167 spectra with H emission features, we subtract 8 known
PCEB candidates and the 4 systems from the previous paragraph, which yields 155 WD-M star binaries, corresponding to a fraction of 24.4\,\%, that probably
harbor an active M dwarf. Although this value matches the number found by \citet{2004AJ....128..426W} for active field M dwarfs in the SDSS, our sample of MS stars mainly consists of stars with spectral types earlier than M5 (see Fig. \ref{fig:comp_RM07-Sil06}). For this range, and in particular for dM stars earlier than M4, the activity fraction from \citet{2004AJ....128..426W} is much smaller than 10\,\%. \citet{2006AJ....131.1674S} detected a significantly higher fraction of active M dwarfs in WD-M binaries (between 20\,\% and 50\,\% from M0 to M5), which they
attributed to a rotational spin-up of the secondary due to
gravitational interaction with the WD. They also reported a lack
of active dM stars with ages $>$ 0.8\,Gyr. Since they imposed a less
stringent ($g-r$) cutoff ($<$ 0.7) for most of their sample than
we had ($<$ 0.2), hotter, and thus younger, WDs passed their
selection process, which would increase their fraction of active
M dwarfs to a higher value than ours. Nevertheless, our results for the fraction of active M dwarfs is in better agreement with the results found by \citet{2006AJ....131.1674S} than it fits to the low fraction found by \citet{2004AJ....128..426W}. Even if we consider only unresolved
binaries, as \citet{2006AJ....131.1674S} did, our fraction of binary systems with
H$_{\alpha}$ emission that probably originates in the M component does not decrease below 23.2\,\%.

\begin{figure}
  \centering
  \scalebox{0.35}{\includegraphics{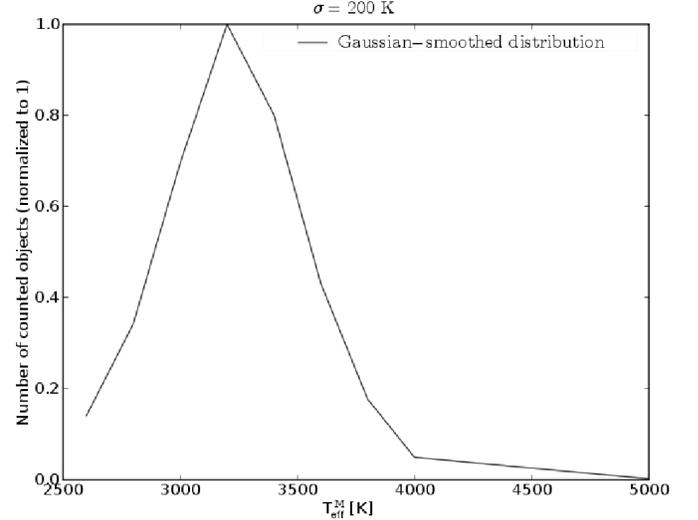}}
  \caption{$T_{\mathrm{eff}}^{\mathrm{M}}$ function; The cutoff at
  2\,600\,K towards lower temperatures is due to our lack of cooler models. The peak at 3\,200\,K is probably generated by SDSS selection effects and does not reflect the true M dwarf population. The tail towards $T_{\mathrm{eff}}^{\mathrm{M}} > 3\,800$\,K is caused by 4 objects with insecure fit at $T_{\mathrm{eff}}^{\mathrm{M}} = 4\,000$\,K and the Gaussian decay.}
 \label{fig:Teff_M}
\end{figure}

\begin{figure}
 \centering
 \scalebox{0.35}{\includegraphics{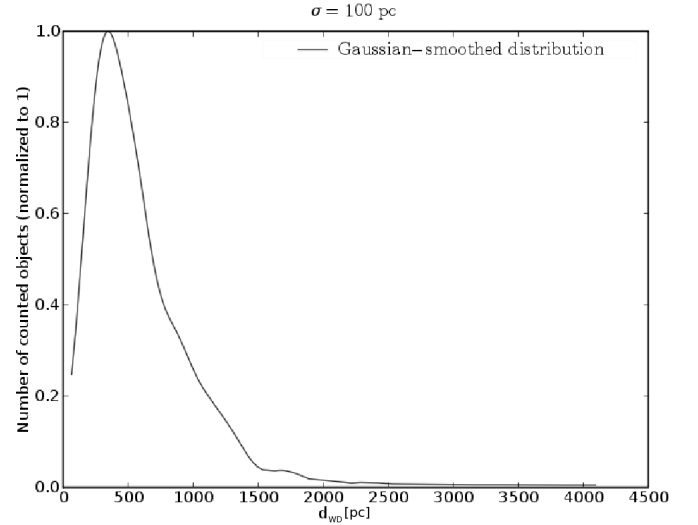}}
 \caption{The distribution of the WD distances in our sample is a consequence of the restricted magnitude range of the SDSS. It has its maximum at 354 pc.}
 \label{fig:dist_WD}
\end{figure}

\begin{figure}
 \centering
 \scalebox{0.35}{\includegraphics{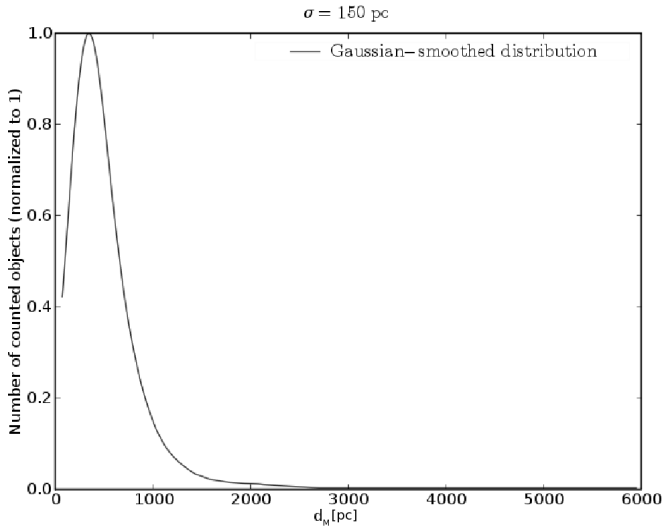}}
 \caption{The distribution of the M star distances in our sample is a consequence of the restricted magnitude range of the SDSS. Its maximum is at 334 pc.}
 \label{fig:dist_M}
\end{figure}

\begin{figure}
 \centering
 \scalebox{0.35}{\includegraphics{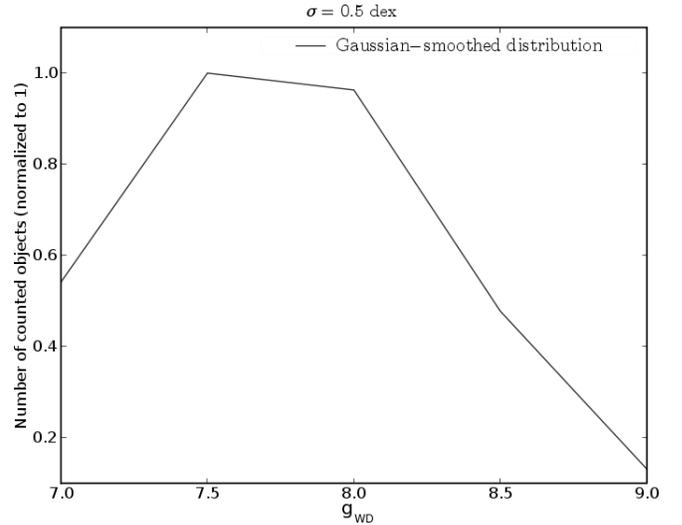}}
 \caption{Distribution of the WD surface gravities. }
 \label{fig:logg_WD}
\end{figure}

\begin{figure}
 \centering
 \scalebox{0.35}{\includegraphics{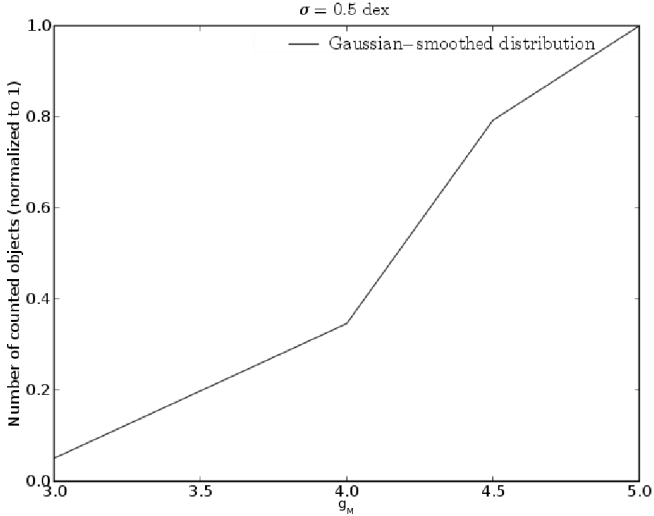}}
 \caption{Distribution of the M star surface gravities; The few objects with $g_{\mathrm{M}} < 4$\,dex are probably M giants.}
 \label{fig:logg_M}
\end{figure}

\begin{figure}
 \centering
 \scalebox{0.35}{\includegraphics{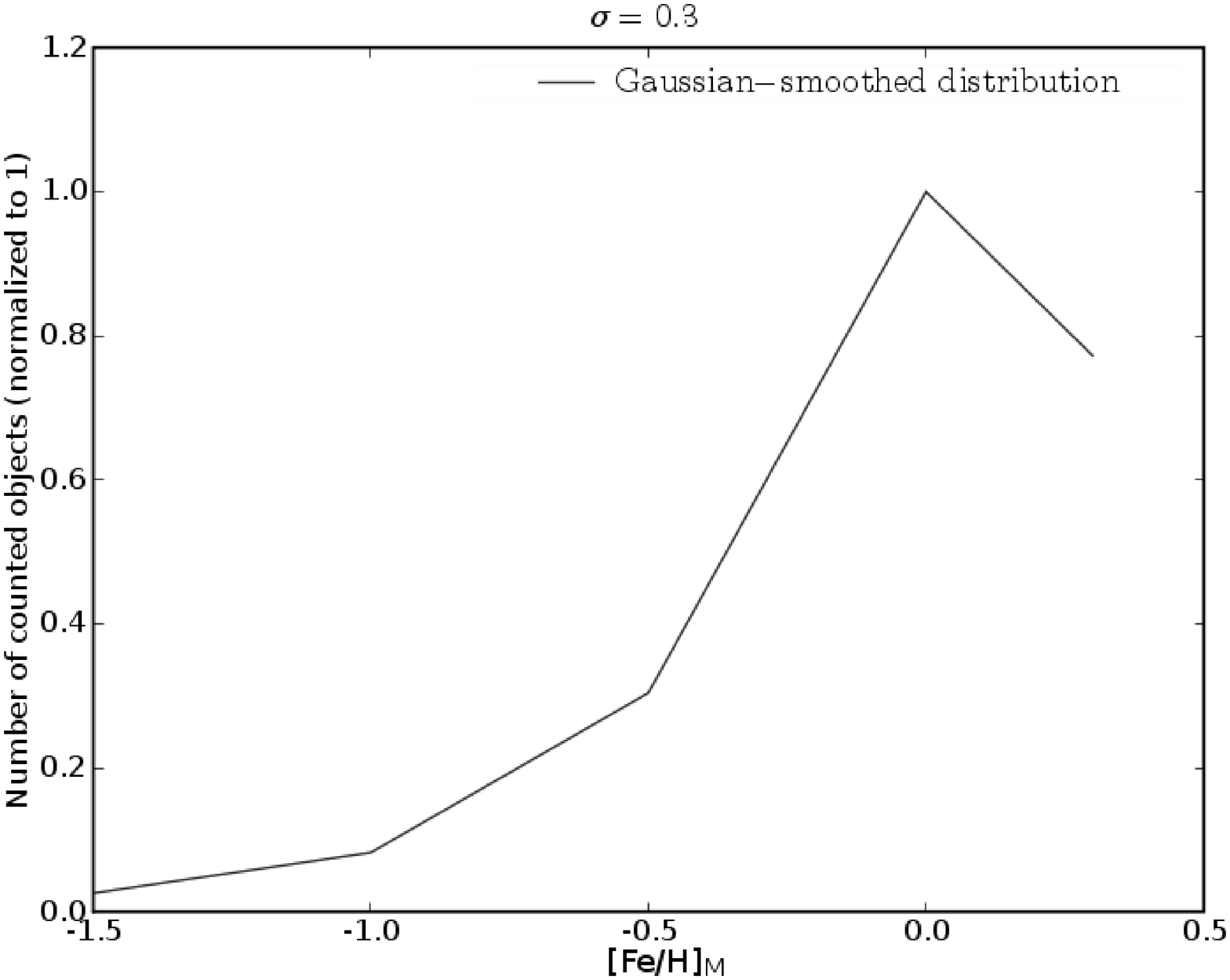}}
 \caption{Metallicities of the M stars in our sample, normalized to 1; The plot shows a distinct peak at [Fe/H]$_\mathrm{M}$ = 0. \texttt{PHOENIX} models for [Fe/H]$_\mathrm{M} > 0.3$ were not available. Among the 70 candidates for a VLMO, there is only J1323+3018 showing a metallicity $>$ 0.}
 \label{fig:Fe-H}
\end{figure}

\subsection{Optically Resolved Binaries}
\label{subsec:optical}

\begin{figure}
 \centering
 \scalebox{0.35}{\includegraphics{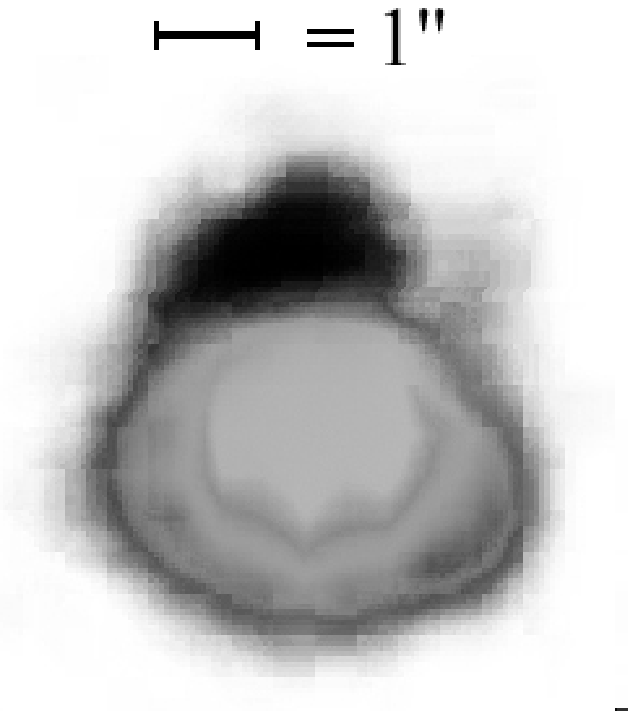}}
 \caption{As an example for one of the 41 optically resolved binaries with further treatment,
          we show the SDSS image of J1127$-$0028. The image resolution is 0.025\,$\arcsec/$pixel.
          In this grayscales image, the red companion is set black while the larger
          and brighter spot is the white dwarf.}
 \label{fig:J1127-0028_OptRes}
\end{figure}

Of our 636 binaries, 41 were chosen for follow-up studies because,
firstly, the red star was found to be located within an area
around the WD that was covered by the SDSS fiber and, secondly,
these stars had a sufficiently wide separation to enable us to
distinguish between the two components. Figure
\ref{fig:J1127-0028_OptRes} provides a typical illustration of
these systems. For photometric binaries with a separation clearly
larger than the SDSS fiber radius of $1.5\,\arcsec$, we assume
that the M star on the SDSS image is not the one represented in
the spectrum. The respective fitting results are indicated in
Table \ref{tab:results} and should be interpreted with caution
since we may be unable to determine if a significant fraction of
the light contribution from the red star was collected by the
fiber, despite the separation being larger than $1.5\,\arcsec$.
The results for the 41 clearly separated objects can be seen in
Table \ref{tab:ResBin}. The typical projected distances are large
and of the order of some hundred AU with a mean value of roughly
650\,AU. The widest separation that we find is $1\,700$\,AU for
J1006+5633, where the true spatial separation should be even
wider. The orbital periods that we derive are typically of the
order of $10^4$\,yr with J1006+5633 indicating a period of
$\gtrsim 70\,000$\,yr.

\subsection{Very Low-Mass Objects}
\label{subsec:vlmo}

We flagged the secondaries in our sample to be very low-mass objects (VLMOs), provided that they had a mass $\leq 0.1\,M_{\sun}$ and $\log(g_{\mathrm{M}}) = 5$, or $M_{\mathrm{M}} \leq 0.09\,M_{\sun}$ and $\log(g_{\mathrm{M}}) \geq 4.5$, to identify potential candidates for substellar companions. Seventy M stars in our reservoir are VLMO according to that definition (see the electronic version of this paper for the table of results). The M masses were calculated from  the fitted $T_{\mathrm{eff}}^{\mathrm{M}}$ and [Fe/H]$_\mathrm{M}$ using evolutionary tables calculated by \citet{1998A&A...337..403B}. Their models are available for $[\mathrm{M/H}] = -2, -1.5, -1.3, -1, -0.5$, and $0$, while no specific tracks exist for those M stars, which are fitted to have $[\mathrm{M/H}] = 0.3$. Since radius and mass at constant $T_{\mathrm{eff}}^{\mathrm{M}}$ decrease monotonously with metallicity in the other tracks, we can only quote the [Fe/H]$_\mathrm{M}$ $= 0$ values as a lower limit for $M_{\mathrm{M}}$ and $R_{\mathrm{M}}$, and consequently also for $d_{\mathrm{M}}$. However, there is only one VLMO candidate in our sample that
has a metallicity of $+0.3$: J1323+3018. The other 69 VLMO candidates are most accurately fitted with $[\mathrm{M/H}] \leq 0$.

If any of the VLMOs were found to be substellar, they would of course by definition be in the cooling phase of their evolution, and more significant deviations from the old main-sequence models would be expected for
these objects, biasing the radius estimate and thus also the assessment of the distance.

\section{Discussion}
\label{sec:discussion}

\begin{figure*}
  \centering
  \scalebox{0.41}{\includegraphics{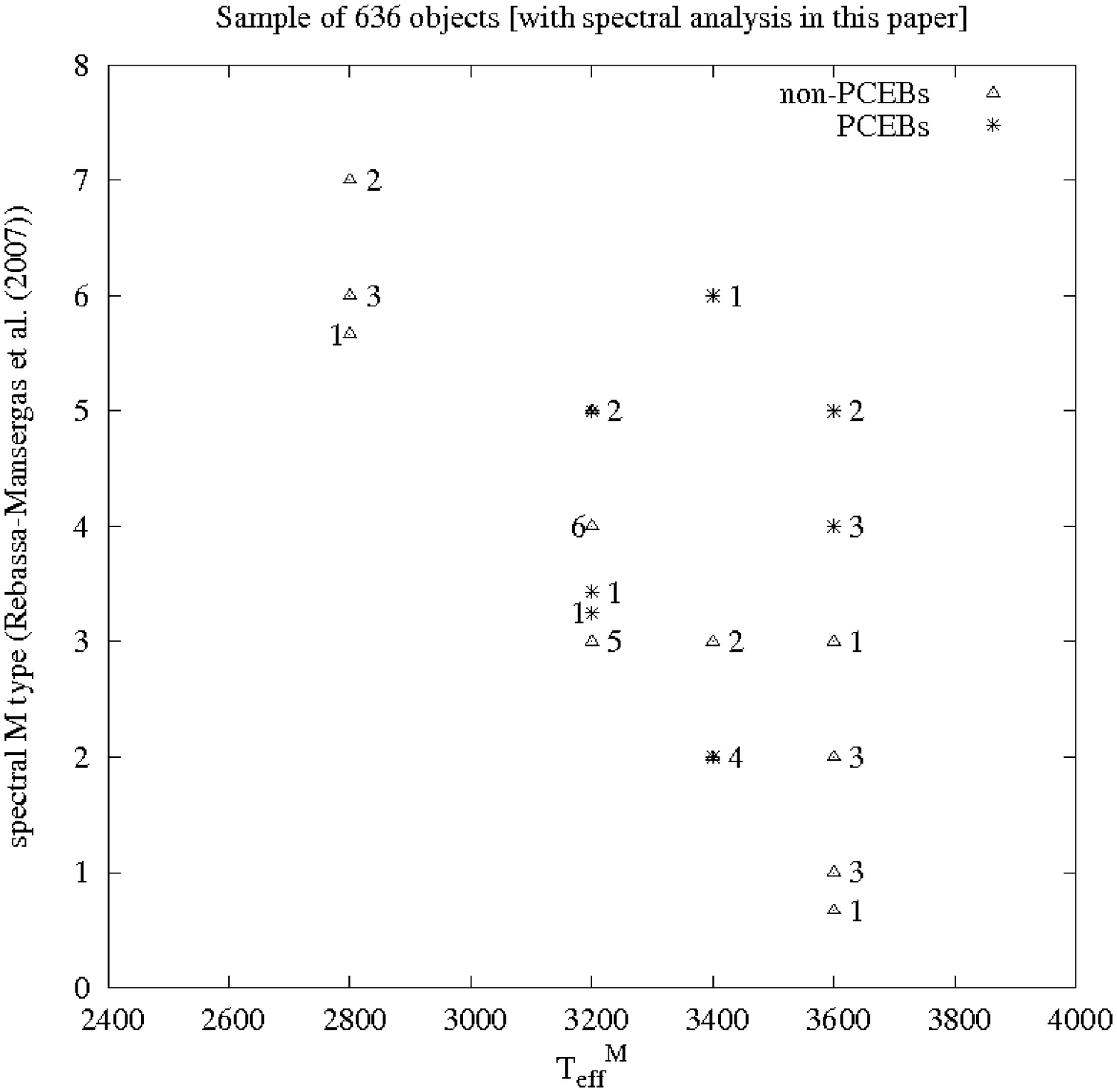}}
  \hspace{1.cm}
  \scalebox{0.41}{\includegraphics{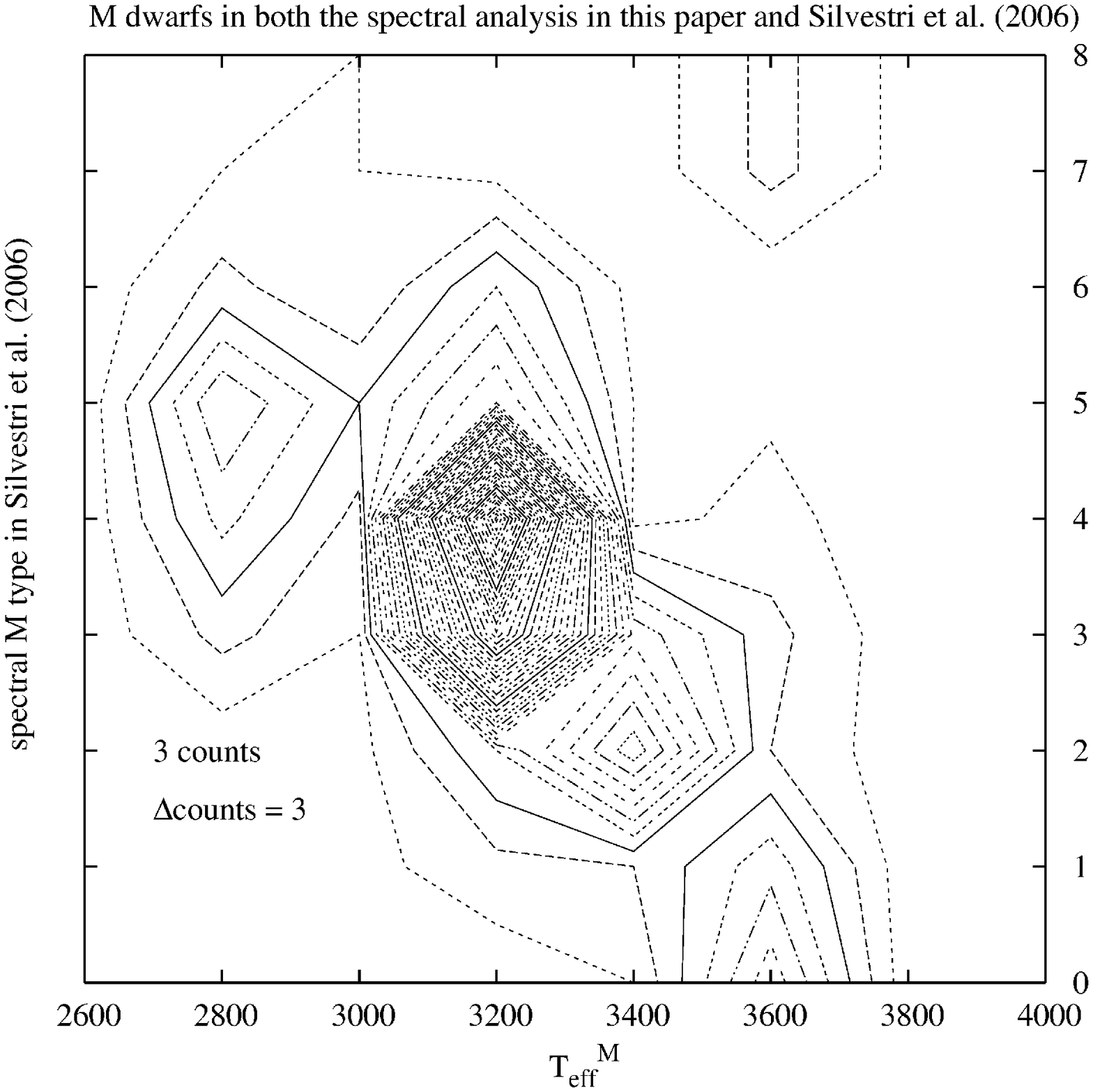}}
  \hspace{1.2cm}
  \caption{\textit{Left}: Comparison of the spectral types given in \citet{2007MNRAS.382.1377R} with the effective temperatures we deduced for 41 M stars, that match both their sample and ours; 11 of them are marked as PCEBs, as confirmed in their study via RV measurements. The labels indicate the counts per grid point. \textit{Right}: Comparison of the spectral types given in \citet{2006AJ....131.1674S} with the effective temperatures we derived for 446 M stars, that match both their sample and ours; The number of M dwarf stars per grid point is projected onto the plane spanned by $T_{\mathrm{eff}}^{\mathrm{M}}$ and the spectral type. The outer contour marks the path of three counts and each subsequent contour symbolizes an increase of three counts. The maximum is at (3\,200\,K\,,\,M4) with 115 counts. Their average spectral mismatch was $\pm$1, while our 1-$\sigma$ accuracy for $T_{\mathrm{eff}}^{\mathrm{M}}$ is 100\,K.}
  \label{fig:comp_RM07-Sil06}
\end{figure*}

The M star temperatures that we fitted are in good agreement with the spectral types derived by \citet{2007MNRAS.382.1377R}. In Fig. \ref{fig:comp_RM07-Sil06}, we compare our fits for $T_{\mathrm{eff}}^{\mathrm{M}}$ with their spectral types for 41 objects (left panel) and with the spectral classification from \citet{2006AJ....131.1674S} (right panel) for 446 objects, which were both in our own and their sample. The authors of the latter publication derived their dM types on the basis of template spectra and color indices as described in \citet{2002AJ....123.3409H}. Since we project the number of M dwarfs per $T_{\mathrm{eff}}^{\mathrm{M}}$ and per spectral type onto the plane spanned by these two parameters, this plot is related to Fig. \ref{fig:Teff_M}. In both of the figures, the maximum is reached at (3\,200\,K\,,\,M4), and the plot shows a monotone decrease of the spectral type with increasing temperature. These counts probably do not relate directly to the true $T_{\mathrm{eff}}^{\mathrm{M}}$ or spectral type function due to selection effects such as the increasing number of M dwarfs towards later types on the one hand, and decreasing visibility of the secondary component on the other hand.

\begin{figure*}
  \centering
  \scalebox{0.42}{\includegraphics{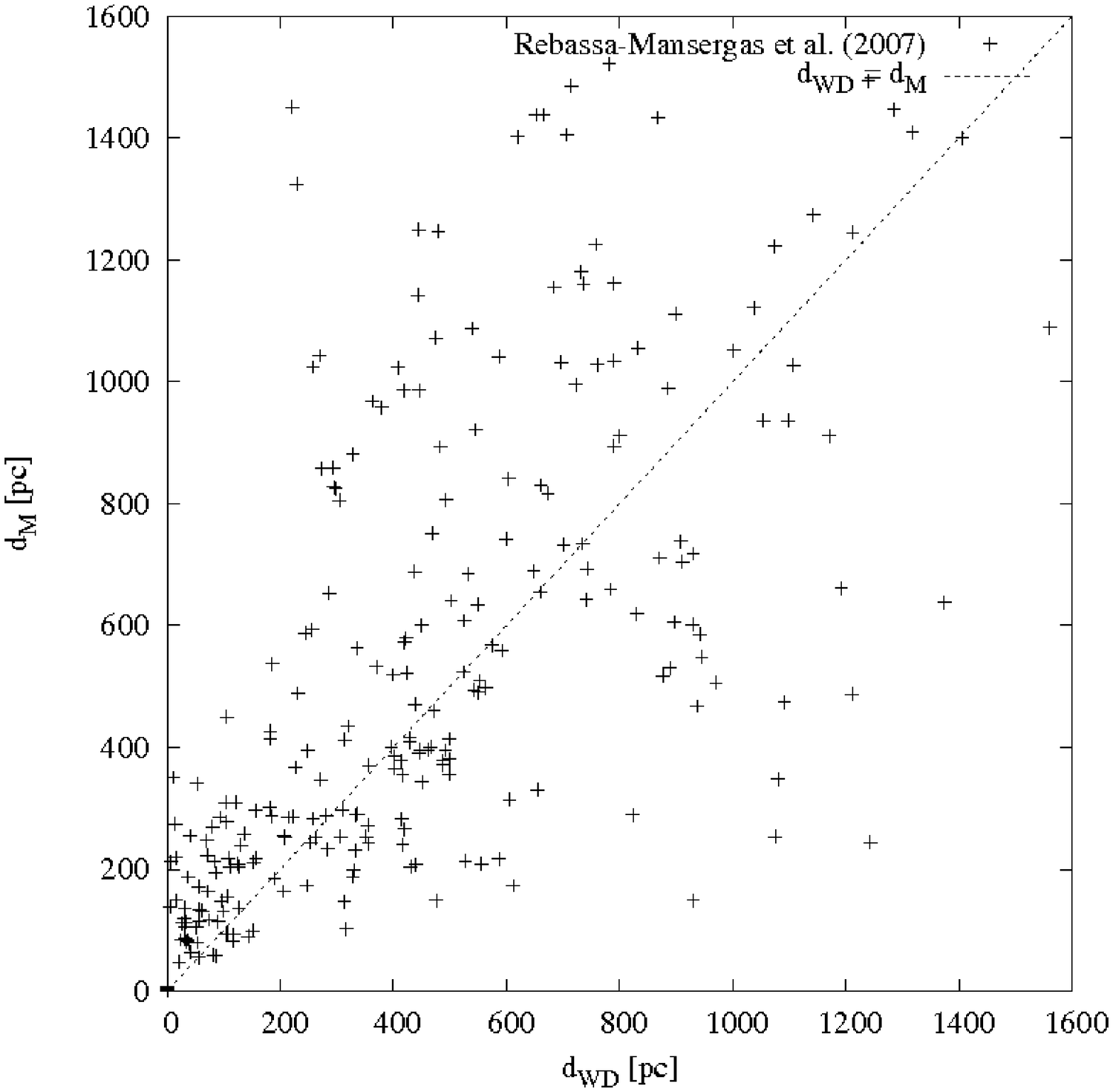}}
  \scalebox{0.45}{\includegraphics{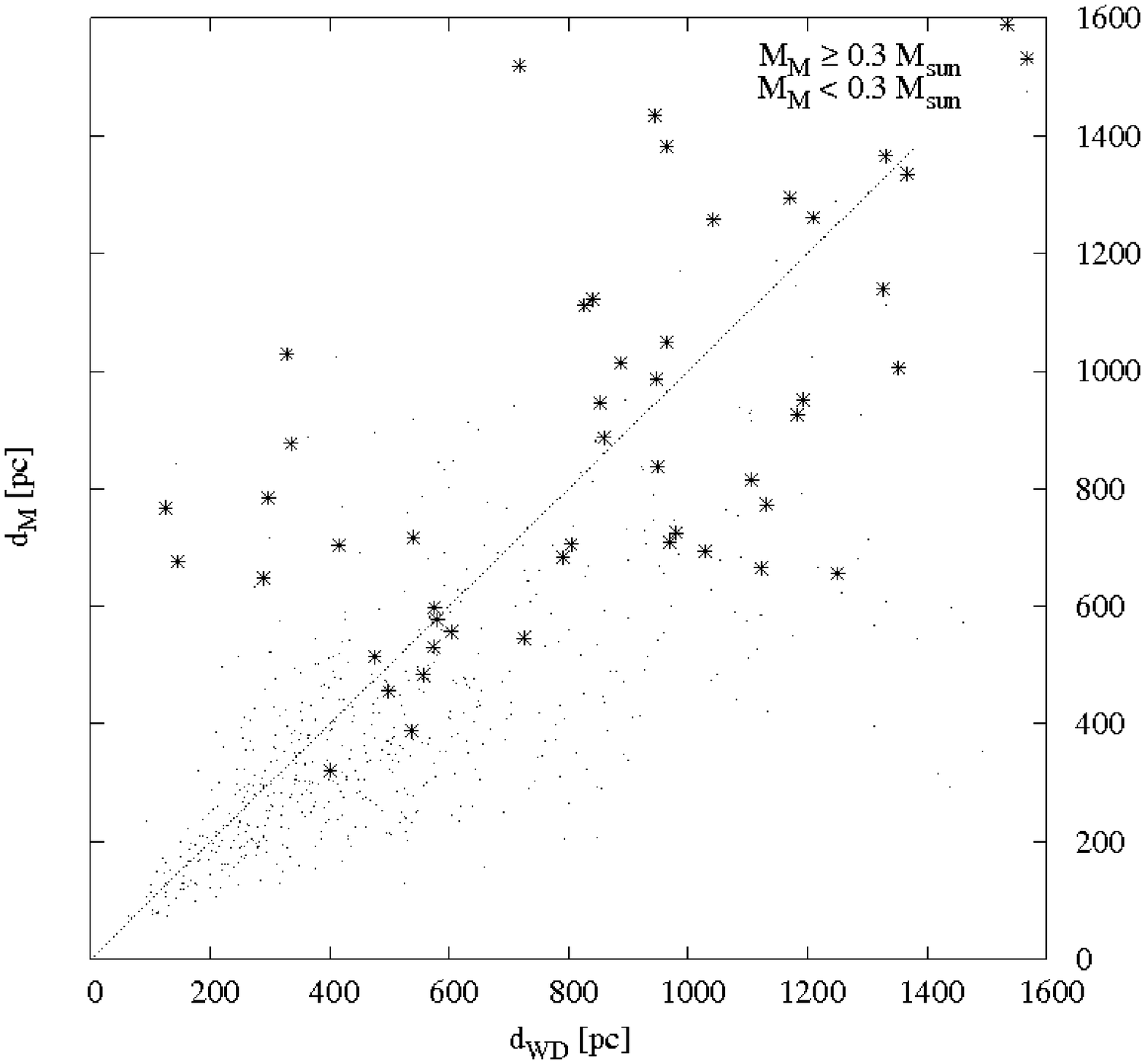}}
  \caption{\textit{Left}: Distribution of the derived distances of the stellar components from \citet{2007MNRAS.382.1377R}; In contrast to our results, their binaries tend to have a larger value for $d_\mathrm{M}$ than for $d_\mathrm{WD}$. \textit{Right}: For M stars with masses $\geq 0.3\,M_{\sun}$ there is no tendency of $d_\mathrm{M} < d_\mathrm{WD}$ in our data. }
  \label{fig:dist_RM07_low-mass}
\end{figure*}

In the \citet{2007MNRAS.382.1377R} study, the MS stars showed a tendency to be at larger distances from Earth than the WD primary (see Fig. \ref{fig:dist_RM07_low-mass}). The authors ascribed this trend to spots on the secondaries' surfaces, which would cause the spectral types to appear too early. The radii derived from the spectral type-radius relation would then have been too large, in addition to the computed distances of the MS stars. In our study, we observe the opposite trend, namely $d_{\mathrm{M}}$ being smaller than $d_{\mathrm{WD}}$, and we consider three possible effects contributing to that trend:

\begin{enumerate}[1.]

\item
\citet{2005AN....326..930R} showed that the absorption in the TiO $\epsilon$-band is systematically underestimated by {\tt PHOENIX} M star spectra. Since as a primary temperature indicator, this band system deepens with decreasing $T_{\mathrm{eff}}^{\mathrm{M}}$, model spectra fits to this feature may also underestimate systematically $T_{\mathrm{eff}}^{\mathrm{M}}$. This effect, if present in our models as well, would cause $d_\mathrm{M}$ to be systematically underestimated. Figure \ref{fig:02} shows indeed that the model predicts a relatively strong TiO $\delta$-band at 8\,870\,{\AA} compared with the $\epsilon$-band at  8\,450\,{\AA}. However, the mismatch is not severe, and the $\gamma$-band at 7\,055\,{\AA} is reproduced well. We therefore do not expect uncertainties in the molecular opacities to introduce a strong bias in our results.

\item
As mentioned in Sect. \ref{sub:premises}, there is observational evidence that the masses of WDs in magnetic, accreting binaries are mainly between 0.7\,$M_{\sun}$ and 0.8\,$M_{\sun}$, instead of around 0.6\,$M_{\sun}$ as for field WDs. There is no evidence for accretion in any of the spectra in our sample. Only 11 out of the 636 objects presented here are known to be close systems, i.e. PCEBs, while 41 are quite widely separated as inferred from the SDSS images. We cannot assess of the remaining orbits on the basis of our data, and we cannot state if the WD masses are close to 0.7 or even 0.8\,$M_{\sun}$. However, a trend towards higher WD masses would be compatible with a bias towards $d_{\mathrm{WD}} > d_{\mathrm{M}}$. An increase in mass leads to a decrease in radius for a WD, and consequently a smaller distance to Earth is required to reproduce the observed flux. Our possible underestimation of $M_{\mathrm{WD}}$ could thus contribute to a systematic overestimation of $d_{\mathrm{WD}}$.

\item
Based on the effective temperature and the metallicity from our spectral fits, the M star radii that we deduced from the \citet{1997A&A...327.1039C} model tracks are systematically underestimated for low-mass objects ($M_\mathrm{M} \lesssim 0.3\,M_{\sun}$, see Fig. \ref{fig:dist_RM07_low-mass}). This discrepancy between theory and observations is well known in the research field of low-mass stars and emerged in observations of eclipsing binary systems with a low-mass component \citep{2007arXiv0711.4451R}. With our -- in statistical terms -- large sample, we support this claim. We encourage the reader to look at Fig. \ref{fig:dist_RM07_low-mass} to see that the trend vanishes for MS stars with masses $0.3\,M_{\sun}$!
\end{enumerate}

The VLMOs in our binary systems (Sect.~\ref{subsec:vlmo}) are
auspicious targets for follow-up investigations of their mass with
time-resolved spectroscopy. Unambiguous mass determinations,
independent from the orbital inclination of the system, are
possible in eclipsing binaries, where the inclination can be taken
from the eclipse light curve or can be assumed to be close to
$90\,^{\circ}$. For systems whose orbits can be measured by
astrometry, the inclination might also be more tightly
constrained; although 4 of the VLMOs presented here are
also optically resolved (see Table \ref{tab:ResBin}),
time-resolved astrometric measurements would be unhelpful since
the orbital periods are of the order of 5\,000\,yr. Furthermore,
even using high-resolution astrometry with adaptive optics, it
would be difficult to resolve systems of less than 10\,yr orbital
periods, since virtually all of our targets are at distances
$>$\,100\,pc.

There is a relatively large number of widely-separated binaries
with orbital distances $\gtrsim$ 250\,AU and up to 1\,700\,AU:
41/636 = 6.4\,\%. This is, however, not a stable assessment; we
suggest five effects that may smear the optical-binary fraction to
either higher or lower values.

\begin{enumerate}[1.]

\item
If two objects are so close together that they appear as one elongated object, the SDSS classification procedure may preselect it as a possible galaxy. This would imply that we have underestimated the true optical-binary fraction.

\item
A small number of physical pairs aligned along the line of sight could also drive the true optical-binary fraction to a higher value. We would have failed them to detect because of their unfavorable geometrical constellation.

\item
Some of the stellar duets on the SDSS images with angular distances $> 1.5\,\arcsec$ could also be physical pairs (but the secondary on the image would not be the one represented in the respective SDSS spectrum).

\item
A significant contingent of binaries with mass overflow from the MS companion to the WD are supposed to be located out of the color-color region that we studied in compiling our sample. These stellar duets are close systems with no optical separation on the SDSS images, which pushes the optical-binary fraction to
lower values.

\item
The statistical considerations presented in Sect. \ref{sub:premises} demonstrated that about 4 of the binaries in our sample should be aligned by chance -- without a common evolutionary background. Probably none, or at most one, of the optical binaries is one of these outliers.

\end{enumerate}

\noindent
To help consider whether stellar duets with such large orbital separations are stable over long timescales, we refer the reader to the paper of \citet{1987ApJ...312..367W} and Fig. 2 therein. Their calculations, including both stars passing by and encounters with subclumps within giant molecular clouds, showed that binaries with a total mass of $1\,M_{\sun}$ and initial orbital separations of around 650\,AU have a typical lifetime of more than the age of the Universe.

\begin{acknowledgements}

R. Heller is supported by a PhD scholarship of the DFG Graduiertenkolleg 1351 ``Extrasolar Planets and their Host Stars''. We thank D.~Koester for his computations of the WD spectra grid and we also acknowledge the anonymous referee for his useful contributions. We also appreciate the remarks by B.~G\"ansicke and S.~Schuh referring to Sect. \ref{subsec:emission}.

Funding for the SDSS and SDSS-II has been provided by the Alfred P. Sloan Foundation, the Participating Institutions, the National Science Foundation, the U.S. Department of Energy, the National Aeronautics and Space Administration, the Japanese Monbukagakusho, the Max Planck Society, and the Higher Education Funding Council for England. The SDSS Web Site is http://www.sdss.org.

The SDSS is managed by the Astrophysical Research Consortium for the Participating Institutions. The Participating Institutions are the American Museum of Natural History, Astrophysical Institute Potsdam, University of Basel, University of Cambridge, Case Western Reserve University, University of Chicago, Drexel University, Fermilab, the Institute for Advanced Study, the Japan Participation Group, Johns Hopkins University, the Joint Institute for Nuclear Astrophysics, the Kavli Institute for Particle Astrophysics and Cosmology, the Korean Scientist Group, the Chinese Academy of Sciences (LAMOST), Los Alamos National Laboratory, the Max-Planck-Institute for Astronomy (MPIA), the Max-Planck-Institute for Astrophysics (MPA), New Mexico State University, Ohio State University, University of Pittsburgh, University of Portsmouth, Princeton University, the United States Naval Observatory, and the University of Washington.

This research has made use of the SIMBAD database, operated at CDS, Strasbourg, France.

\end{acknowledgements}

\bibliographystyle{aa} 
\bibliography{0632_Heller_et_al_WD-M_2008}

\begin{table*}[h!]
  \centering
  \caption[Optically resolved binaries]{Optically resolved binaries (41 objects); We use the mathematically exact values of $d_{\mathrm{proj}}$ and $d_{\mathrm{proj}}^{\mathrm{corr}}$ to compute the lower limit for $P$ and the statistically corrected $P^{\mathrm{corr}}$. But for this table, distances greater than 500\,pc or 500\,AU are rounded to the next hundred and else to the next fifty. Periods are rounded to the next 500\,yr for $P < 10\,000$\,yr, to the next 1\,000\,yr for 10\,000\,yr $< P <$ 50\,000\,yr and else to the next 5\,000\,yr. $^{\dag}$: VLMO candidates.}
  \label{tab:ResBin}
    \begin{tabular}{lcrrrcrr}

    \hline

    \hline

    \hline \hline

    Designation & $d_{\mathrm{proj}}$\,[$\arcsec$] & $\overline{d}$ [pc] & $d_{\mathrm{proj}}$ [AU] & $d_{\mathrm{proj}}^{\mathrm{corr}}$\,[AU] & $M_{\mathrm{M}}$\,[$M_{\sun}$] & $P$\,[yr] $\gtrsim$ & $P^{\mathrm{corr}}$\,[yr]\\
    \hline

     J0017$-$0009          & 1.4 &        500 &        700 &   1\,000 & 0.46 &        16\,000 &  33\,000 \\
     J0122+1542            & 1.0 &        300 &        300 &      500 & 0.17 &         6\,500 &  13\,000 \\
     J0151$-$0800          & 1.0 &        350 &        350 &      600 & 0.17 &         8\,000 &  16\,000 \\
     J0215+1418            & 0.9 &        700 &        600 &      900 & 0.27 &        16\,000 &  30\,000 \\
     J0249+3342            & 1.4 &        500 &        700 &   1\,200 & 0.17 &        23\,000 &  45\,000 \\
     J0348$-$0614          & 1.1 &        500 &        600 &      900 & 0.17 &        16\,000 &  31\,000 \\
     J0725+4145            & 1.3 &        400 &        500 &      900 & 0.17 &        15\,000 &  29\,000 \\
     J0729+4304            & 1.5 &        150 &        250 &      400 & 0.12 &         5\,000 &  10\,000 \\
     J0739+2743            & 0.6 &        700 &        350 &      600 & 0.17 &         8\,000 &  16\,000 \\
     J0740+3859            & 1.1 &        300 &        350 &      600 & 0.12 &         8\,000 &  15\,000 \\
     J0741+3808            & 0.9 &        300 &        250 &      400 & 0.17 &         5\,000 &   9\,000 \\
     J0752+4332            & 1.3 &        600 &        800 &   1\,200 & 0.18 &        23\,000 &  46\,000 \\
     J0800+5002            & 1.1 &        700 &        800 &   1\,200 & 0.27 &        22\,000 &  44\,000 \\
     J0801+2216            & 1.3 &        500 &        700 &   1\,100 & 0.17 &        20\,000 &  40\,000 \\
     J0806+4035            & 1.2 &        350 &        450 &      700 & 0.17 &        11\,000 &  21\,000 \\
     J0809+1251            & 1.3 &        400 &        500 &      800 & 0.17 &        14\,000 &  27\,000 \\
     J0813+2152            & 0.9 &        900 &        800 &   1\,300 & 0.18 &        28\,000 &  55\,000 \\
     J0829+2701            & 1.4 &     1\,000 &     1\,400 &   2\,200 & 0.27 &        55\,000 & 110\,000 \\
     J0845+2348            & 1.1 &     1\,400 &     1\,400 &   2\,200 & 0.46 &        50\,000 & 100\,000 \\
     J0904+5621$^{\dag}$   & 1.5 &        200 &        300 &      450 & 0.10 &         6\,000 &  12\,000 \\
     J0931+3941            & 1.4 &        350 &        500 &      800 & 0.17 &        13\,000 &  26\,000 \\
     J0939+5729            & 1.3 &        250 &        300 &      500 & 0.17 &         6\,000 &  12\,000 \\
     J0942+1846            & 1.3 &        600 &        700 &   1\,200 & 0.17 &        23\,000 &  45\,000 \\
     J1001+3203            & 1.3 &        250 &        350 &      500 & 0.17 &         7\,000 &  13\,000 \\
     J1006+5633            & 1.4 &     1\,200 &     1\,700 &   2\,700 & 0.46 &        70\,000 & 135\,000 \\
     J1032+3722            & 1.3 &        350 &        450 &      700 & 0.17 &        10\,000 &  20\,000 \\
     J1127$-$0028$^{\dag}$ & 1.5 &        200 &        250 &      400 & 0.10 &         4\,500 &   9\,000 \\
     J1127+4249            & 1.2 &        350 &        400 &      600 & 0.12 &         9\,000 &  18\,000 \\
     J1205+0312            & 1.4 &        250 &        400 &      600 & 0.17 &         8\,500 &  17\,000 \\
     J1209+6510$^{\dag}$   & 1.4 &        250 &        350 &      500 & 0.10 &         7\,500 &  14\,000 \\
     J1210+0549            & 1.4 &        350 &        500 &      800 & 0.17 &        12\,000 &  24\,000 \\
     J1216+4328            & 1.5 &        900 &     1\,300 &   2\,000 & 0.46 &        46\,000 &  90\,000 \\
     J1242+4506            & 1.5 &        450 &        700 &   1\,100 & 0.17 &        20\,000 &  40\,000 \\
     J1253+5813            & 1.4 &        400 &        600 &      900 & 0.27 &        14\,000 &  28\,000 \\
     J1304+1449            & 1.3 &        500 &        600 &   1\,000 & 0.17 &        18\,000 &  35\,000 \\
     J1347+4129            & 1.3 &     1\,200 &     1\,600 &   2\,500 & 0.46 &        65\,000 & 125\,000 \\
     J1456+4824            & 1.4 &        450 &        600 &   1\,000 & 0.17 &        18\,000 &  35\,000 \\
     J1606+4217            & 1.1 &        500 &        600 &      900 & 0.17 &        17\,000 &  35\,000 \\
     J1630+1302            & 1.4 &        600 &        900 &   1\,420 & 0.17 &        31\,000 &  60\,000 \\
     J1744+2442            & 1.4 &     1\,100 &     1\,500 &   2\,400 & 0.46 &        55\,000 & 115\,000 \\
     J2200$-$0715$^{\dag}$ & 1.5 &        200 &        250 &      400 & 0.10 &         5\,000 &  10\,000 \\

    \hline
  \end{tabular}
\end{table*}

\begin{landscape}

\thispagestyle{empty}
\topmargin 0.4cm

\begin{table}
  \caption[Example table of results for 636 WD-M star binaries]{Example table of results for 636 WD-M star binaries; Typical uncertainties are mentioned in the text. The 11 PCEBs and PCEB candidates from \citet{2007MNRAS.382.1377R}, as well as the two CV candidates mentioned in Sect. \ref{subsec:emission}, are shaded in gray. $^*$: M component probably a giant due to the very low surface gravity of 3\,dex. $^{\dag}$: VLMO candidates. \textit{H}: H$_\alpha$ and other Balmer line emission. rm*: WD parameters taken from \citet{2007MNRAS.382.1377R}. $\bullet$: Clearly resolved objects with projected angular distance $< 1.5\,\arcsec$. $\circ$: Objects with slight angular separation and without further treatment. $\star$: Objects separated by more than $1.5\,\arcsec$, likely to be not those represented in the spectrum. ?: The photometric data corresponding to the object id specified does not exist in the SDSS database. The complete table is available as Supplementary Material to the online version of this article at the CDS via anonymous ftp to cdsarc.u-strasbg.fr (130.79.128.5) or via http://cdsweb.u-strasbg.fr/cgi-bin/qcat?J/A+A.}
  \label{tab:results}
  \begin{tabular}{l|rrrrrcrcccrrccccccr}

    \hline

    \hline

    \hline \hline

    Designation & $d_\mathrm{WD}$ [pc] & $d_\mathrm{M}$ [pc] & $\mathcal{C}$ & $T_\mathrm{eff}^\mathrm{WD}$ [K] & $T_\mathrm{eff}^\mathrm{M}$ [K] & $R_\mathrm{WD} [R_{\sun}]$ & $R_\mathrm{M} [R_{\sun}]$ & log $g_\mathrm{WD}$ & log $g_\mathrm{M}$ & $M_\mathrm{WD} [M_{\sun}]$ & $M_\mathrm{M} [M_{\sun}]$ & [Fe/H]$_\mathrm{M}$ & Emission & WD & Res. & $m$ & $N$ & $\chi^2_\mathrm{red,min}$\\

    \hline

      J0017$-$0009 & 475.8 & 514.6 & $-$0.055 & 50000.0 & 3600.0 & 0.016 & 0.433 & 7.5 & 4.5 & 0.6 & 0.46 & $-$0.0 & \textit{H} & DA & $\bullet$ & 3783 & 1572 & 4.29\\ 
      J0017+0040 & 547.2 & 253.5 & 0.518 & 13000.0 & 3200.0 & 0.013 & 0.189 & 7.5 & 5.0 & 0.6 & 0.17 & $-$0.0 &  & DA & $\star$ & 3824 & 1599 & 4.10\\ 
      J0039+2548 & 370.9 & 335.9 & 0.070 & 34000.0 & 3200.0 & 0.015 & 0.189 & 8.0 & 5.0 & 0.6 & 0.17 & $-$0.0 &  & DA &  & 3838 & 1599 & 2.26\\ 
      J0041+1511A$^*$ & 1332.0 & 607.1 & 0.529 & 32000.0 & 3600.0 & 0.014 & 0.190 & 7.5 & 3.0 & 0.6 & 0.18 & $-$1.0 &  & DA &  & 3842 & 1602 & 2.74\\ 
\textcolor{Gray}{J0052$-$0053}$^*$ & 579.9 & $>$ 577.2 & $<$ 0.003 & 15000.0 & 3600.0 & 0.008 & $>$ 0.433 & 8.5 & 3.0 & 1.0 & $>$ 0.46 & +0.3 & \textit{H}, rm*& DA &  & 3843 & 1596 & 19.69\\ 
\textcolor{Gray}{J0054$-$0025} & 385.1 & $>$ 383.1 & $<$ 0.003 & 17000.0 & 3200.0 & 0.015 & $>$  0.189 & 8.0 & 4.0 & 0.5 & $>$ 0.17 & +0.3 & \textit{H}, rm*& DA &  & 3843 & 1596 & 6.51\\ 

    \hline
  \end{tabular}
\end{table}

\begin{table}
\flushleft
\caption{Excerpt of our catalog, the master sample, of 857 WD-M binary stars. s: \citet{2006AJ....131.1674S}, r: \citet{2003AJ....125.2621R}, kl: \citet{2004ApJ...607..426K}, rm: \citet{2007MNRAS.382.1377R} (non-PCEBs), rm*: \citet{2007MNRAS.382.1377R} (PCEB), e: \citet{2006ApJS..167...40E}, h: \citet{2006A&A...454..617H}, g: \citet{1986AJ.....92..867G}, x: There was no alternative name found. The complete table is available as Supplementary Material to the online version of this article at the CDS via anonymous ftp to cdsarc.u-strasbg.fr (130.79.128.5) or via http://cdsweb.u-strasbg.fr/cgi-bin/qcat?J/A+A.}
  \label{tab:master_sample}
  \begin{tabular}{lllll}

    \hline

    \hline

    \hline \hline

    Short Name (SDSS) & Full Name (SDSS) & Spectrum File & References & alt.\ Names (as found in the SIMBAD database)\\

    \hline

    J0001$+$0006 & J000152.10$+$000644.7 & spSpec-51791-0387-157.fit &                & x        \\
J0017$+$0040 & J001733.59$+$004030.4 & spSpec-51795-0389-614.fit &    s6, s7,r,kl &          SDSS J001733.59+004030.4 (00\farcs0) \\
J0017$-$0009 & J001749.25$-$000955.4 & spSpec-51795-0389-112.fit & s6, s7,r,kl,rm &          SDSS J001749.25-000955.4 (00\farcs1) PB  5848 (11\farcs5) \\
J0026$+$1444 & J002620.41$+$144409.5 & spSpec-52233-0753-079.fit &         s6, s7 & PHL 2888 \\
J0036$+$0700 & J003602.59$+$070047.3 & spSpec-53709-2312-164.fit &                & PB 6052  PB  6052 (08\farcs1) \\
J0039$+$2548 & J003925.22$+$254823.7 & spSpec-53327-2038-380.fit &                & x        \\

    \hline

  \end{tabular}
\end{table}

\end{landscape}

\end{document}